%% file: main.tex
\documentclass[
reprint,
superscriptaddress,
]{revtex4-2}

\usepackage[
  colorlinks=true,
  urlcolor=blue,
  linkcolor=blue,
  citecolor=blue
]{hyperref}

\usepackage{amsmath}
\usepackage{physics}
\usepackage{lipsum}
\usepackage[euler]{textgreek}
\usepackage[utf8]{inputenc}
\usepackage[margin=0.75in]{geometry}
\usepackage{graphicx}
\usepackage{bbold}
\usepackage{xcolor}
\usepackage{diagbox}
\usepackage{adjustbox}
\usepackage{makecell}
\usepackage{multirow}
\usepackage{tabularx}
\usepackage{enumitem}
\usepackage{cleveref}

\begin{document}
\title{High-fidelity neutral atom gates leveraging low-rank Hessian optimization}

\author{Genyue~Liu}
\thanks{These authors contributed equally to this work.}
\affiliation{Princeton University, Department of Electrical and Computer Engineering, Princeton, New Jersey 08544}

\author{Guillaume~Bornet}
\thanks{These authors contributed equally to this work.}
\affiliation{Princeton University, Department of Electrical and Computer Engineering, Princeton, New Jersey 08544}

\author{Deniz~Kurdak}
\affiliation{Princeton University, Department of Electrical and Computer Engineering, Princeton, New Jersey 08544}

\author{Mingxuan~Xiao}
\affiliation{Princeton University, Department of Electrical and Computer Engineering, Princeton, New Jersey 08544}

\author{Chenyuan~Li}
\affiliation{Princeton University, Department of Physics, Princeton, New Jersey 08544}

\author{Bichen~Zhang}
\affiliation{Princeton University, Department of Electrical and Computer Engineering, Princeton, New Jersey 08544}

\author{Jeff~D.~Thompson}
\thanks{jdthompson@princeton.edu}
\affiliation{Princeton University, Department of Electrical and Computer Engineering, Princeton, New Jersey 08544}

\begin{abstract}
    Quantum optimal control can produce fast and robust multi-qubit gates, but experimentally calibrating the resulting high-dimensional waveforms remains challenging because direct searches over large parameter spaces converge slowly. Building on the low-rank structure of quantum-control landscapes, we develop and benchmark a Hessian-based calibration method for optimal-control gates. The method identifies the few waveform directions that affect fidelity to leading order, with the number of directions set by the accessible leakage and coherent error channels, and optimizes only within this principal space using closed-loop experimental feedback. We apply this approach to an amplitude-robust controlled-$Z$ gate on metastable-state $^{171}$Yb nuclear-spin qubits. Experimentally, we verify the predicted Hessian-sensitive directions and demonstrate rapid convergence of the optimization protocol. The optimized gate reaches a raw fidelity of 0.9959(2), increasing to 0.99902(7) after postselection on no detected loss, and the performance is essentially unchanged under laser-power variations of up to 20\%. We further show that the same fidelity Hessian directions can correct certain Hamiltonian parameter errors. These results establish low-rank Hessian optimization as an efficient and physically motivated calibration strategy for high-dimensional optimal-control gates, which is broadly applicable to many qubit types. 
\end{abstract}

\maketitle


High-fidelity gates are a central requirement for quantum computing and especially for fault-tolerant architectures, where physical gate errors strongly affect the overhead needed for error correction. Quantum optimal control is an important tool for pushing gate performance beyond simple analytic pulses~\cite{palao_quantum_2002, khaneja_optimal_2005, caneva_chopped_2011, machnes_tunable_2018, wilhelm_introduction_2020}. By shaping time-dependent control fields, optimal control can produce gates that are faster, more robust, and better adapted to realistic experimental constraints~\cite{khaneja_optimal_2005, sporl_optimal_2007, choi_optimal_2014, leung_robust_2018, kang_batch_2021, levine_parallel_2019, jandura_time-optimal_2022, fromonteil_protocols_2023, jandura_optimizing_2023, glaser_closed-loop_2025}.

However, realizing this benefit in practice is challenging: the pulses are only as accurate as the model Hamiltonian used to optimize them, and can only be implemented as faithfully as the transfer function of the devices generating the physical control fields. The power of optimal control comes from using an expressive, high-dimensional basis set of control waveforms, but experimentally optimizing a high-dimensional space is prohibitive. A number of optimization approaches have been explored, including analytically designed pulses~\cite{levine_parallel_2019}, parameterized ansatzes~\cite{theis_high-fidelity_2016, evered_high-fidelity_2023, muniz_high-fidelity_2025}, expansion in a truncated basis set~\cite{kelly_optimal_2014, ma_high-fidelity_2023, werninghaus_leakage_2021}, machine learning~\cite{sivak_model-free_2022, porotti_gradient-ascent_2023}, and genetic algorithms~\cite{white_extracting_2004}. However, these approaches have shortcomings, including requiring many experimental samples, not converging to the optimal fidelity, or depending sensitively on independent characterizations of the Hamiltonian or transfer function.

Pioneering theoretical work in quantum optimal control has established that the fidelity landscape around any optimum is robust, in the sense that the fidelity is insensitive to most perturbations of the gate itself~\cite{rabitz_topology_2006, shen_quantum_2006, ho_landscape_2009}. This insight was recently used to identify a reduced-dimensional parameter space for experimental optimization of single-qubit gates in a superconducting processor~\cite{berger_dimensionality_2024}. However, the application of these techniques to optimizing multi-qubit gates, model Hamiltonian errors, and leakage errors is an open challenge.

In this work, we theoretically develop and experimentally benchmark a rigorous model of the sensitivity of multi-qubit gates to control and Hamiltonian errors. We provide an explicit formula for the rank of the space of control waveform perturbations that affect the gate fidelity, expressed in terms of the number of error channels in the target unitary, including leakage to non-computational states. We argue that experimental optimization of the waveform within this restricted space is both necessary and sufficient to correct small but arbitrary errors in the control waveform and many errors in the model Hamiltonian. In the context of two-qubit entangling gates on neutral atom qubits, the relevant control space rank is only 5 or 10 depending on the number of Rydberg levels involved. We apply this to optimize the fidelity of an amplitude-robust controlled-Z (CZ) gate in metastable $^{171}$Yb qubits, reaching an ultimate fidelity of $\mathcal{F}=0.9959(2)$ [$\mathcal{F}_\mathrm{ps}=0.99902(7)$ postselected on survival]. These results establish low-rank Hessian optimization as an efficient and physically motivated method for calibrating optimal-control gates, with potential applications beyond neutral-atom processors.

\section{Low-rank Hessian Optimization}

\begin{figure*}[ht]
    \centering
    \includegraphics[width=\textwidth]{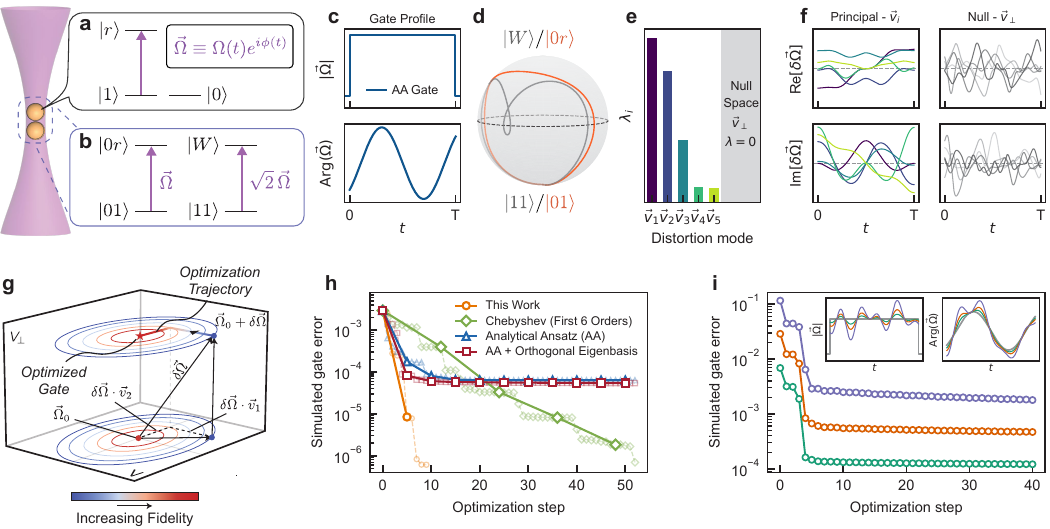}
    \caption{
    \textbf{Symmetric CZ gates and low-rank Hessian optimization.}
    (a) Minimal three-level system for a symmetric Rydberg-mediated CZ gate, in which a pulse couples the qubit state $\ket{1}$ to a Rydberg state $\ket{r}$ with complex Rabi frequency $\vec{\Omega}\equiv\Omega(t)e^{i\phi(t)}$.
    (b) In the two-qubit basis, the $\ket{01}$ and $\ket{10}$ states couple to the Rydberg manifold with Rabi frequency $\vec{\Omega}$, while the $\ket{11}$ state couples to the symmetric state $\ket{W}=(\ket{1r} + \ket{r1})/\sqrt{2}$ with enhanced strength $\sqrt{2}\,\vec{\Omega}$.
    (c) Amplitude and phase profiles of the analytical ansatz (AA) gate, used here as a representative example for studying pulse-distortion sensitivity. (d) The corresponding Bloch sphere trajectories of the initial states $\ket{11}$ and $\ket{01}$.
    (e) Eigenvalue spectrum of the fidelity Hessian of the AA pulse. Note that only five eigenvalues are nonzero.
    (f) Eigenvectors $\vec{v}_i$ of the principal space plotted in their real and imaginary components, together with several null-space modes, $\vec{v}_{\perp}$.
    (g) The optimization procedure corrects the projection of the gate error onto the principal space, $\delta\vec{\Omega}_{\parallel}$, while ignoring the projection onto the fidelity Hessian null space.
    (h) Simulated comparison of optimization strategies (see text). Each optimization cycle (solid markers) consists of multiple optimization steps (lighter markers), with the number of steps depending on the chosen optimization strategy.
    (i) Optimization of distorted initial pulses with three different distortion amplitudes with the AA parametrization. Insets show the corresponding changes to the amplitude and phase of $\vec{\Omega}$.
    }
    \label{fig:Fig1}
\end{figure*}

We develop the concept of low-rank Hessian optimization in the context of entangling gates for neutral atom qubits based on the Rydberg blockade~\cite{Jaksch.2000, isenhower2010demonstration, wilk2010entanglement}. The most common implementation is the symmetric CZ gate, where a single laser field applied to a pair of atoms couples one of the qubit levels in each atom to a Rydberg state~\cite{levine_parallel_2019, saffman2020symmetric}. If the atoms are sufficiently close, the van der Waals interaction prevents simultaneously exciting both atoms to the Rydberg state, resulting in the constrained dynamics shown in Fig. \ref{fig:Fig1}a and b. This can give rise to a controlled-Z gate through the appropriate choice of drive waveform $\Omega = \Omega(t) e^{i \phi(t)}$. Numerous analytic~\cite{levine_parallel_2019, saffman2020symmetric, pagano_error_2022, fromonteil_protocols_2023, evered_high-fidelity_2023} and numerical~\cite{jandura_time-optimal_2022, jandura_optimizing_2023} variants of this gate have been proposed.

In practice, both analytic and numerical gates are sensitive to imperfections in the control waveform, or errors in the model used to derive the gate, necessitating closed-loop experimental calibration. However, identifying intuitive calibration parameters is challenging.
Here, we demonstrate that the eigenvectors of the fidelity Hessian provide a natural, low-rank basis set for closed-loop experimental optimization of arbitrary gates. To analyze the eigenvalue structure, we decompose the gate waveform in an arbitrary orthonormal set of real-valued basis functions $\{f_k(t)\}$ with dimension $N \gg 1$,
\begin{equation}
    \Omega(t)e^{i\phi(t)}
    \approx
    \sum_{k=1}^{N}(\alpha_{2k-1}+i\alpha_{2k})f_k(t).
\end{equation}
The real coefficients $\alpha_i$ define a control vector $\vec{\Omega}=\{\alpha_1,\dots,\alpha_{2N}\}$. We denote the intended gate by a control vector $\vec{\Omega}_0$, which implements the target unitary optimally with fidelity $\mathcal{F} = 1$. Given an erroneous gate $\vec{\Omega}_0+\delta\vec{\Omega}$ (\emph{i.e.}, from distortion in the control waveform or errors in the model Hamiltonian used to compute $\vec{\Omega}_0$), the gate infidelity can be expressed as:
\begin{equation}
    1-\mathcal{F}
    =
    \frac{1}{2}
    \delta\vec{\Omega}^{T}\,
    \mathcal{H}\,
    \delta\vec{\Omega}
    +
    \mathcal{O}(|\delta\vec{\Omega}|^3).
\end{equation}
Here $\mathcal{H}_{ij} = -\frac{\partial^2 \mathcal{F} }{\partial \alpha_i \partial\alpha_j}$ is the Hessian of the gate error with respect to the real control parameter $\alpha_i$, evaluated at the designed waveform. Because $\mathcal{H}$ is a real symmetric matrix, diagonalizing it gives orthonormal eigenvectors $\vec{v}_i$ with corresponding eigenvalues $\lambda_i$. The eigenvectors identify the waveform distortions to which the gate is sensitive, while the eigenvalues quantify the corresponding sensitivities. Thus, the error reads:
\begin{equation}
    1-\mathcal{F}
    = \frac{1}{2}\sum_{i}\lambda_i|\vec{v}_i \cdot \delta\vec{\Omega}|^2.
    \label{eq:error_Hessian_basis}
\end{equation}

Evaluating this function for a representative Rydberg CZ gate~\cite{evered_high-fidelity_2023, muniz_high-fidelity_2025} (Fig.~\ref{fig:Fig1}c and d), we find that there are only five nonzero eigenvalues, even though the control waveform dimension $N$ is very large (Fig.~\ref{fig:Fig1}e). The associated eigenvectors (Fig.~\ref{fig:Fig1}f) form a subspace $V = \mathrm{span}\{\vec{v}_1,\ldots,\vec{v}_r\}$, which we call the principal space. Any waveform distortion can be decomposed into a component inside $V$ and a component perpendicular to it (Fig.~\ref{fig:Fig1}g):
\begin{equation}
    \delta\vec{\Omega}
    =
    \delta\vec{\Omega}_{\parallel}
    +
    \delta\vec{\Omega}_{\perp}.
\end{equation}
The perpendicular component lies in the null space of the Hessian and therefore does not contribute errors to leading order.

The principal space $V$ is independent of the choice of basis functions $\{f_k\}$; it is instead a property of the gate itself, and can be understood from considering the space of accessible errors in the target unitary. In the perturbative regime, a waveform distortion $\delta\vec{\Omega}$ induces only a small change $\delta U$ in the final unitary relative to the ideal gate $U_0$. The map from $\delta\vec{\Omega}$ to $\delta U$ can then be approximated by a linear transformation, where $\delta U$ can be viewed as a tangent vector to $SU(D)$, where $D$ is the dimension of the Hilbert space for the gate. Since this tangent space has dimension $D^2-1$, the dimensionality of the principal space is also bounded to $\dim V \leq D^2 - 1$~\cite{ho_landscape_2009, berger_dimensionality_2024}.

In practice, the relevant rank can be much smaller, because symmetries in the control Hamiltonian restrict the set of possible perturbations $\delta U$. The maximum possible Hessian rank for a gate on the Hilbert space $\{\ket{0},\ket{1},\ket{r}\}^{\otimes 2}$ is 80; however, the smaller rank of five can be explained by a simple formula counting the number of accessible leakage channels and phase errors (Appendix~\ref{sec:rank_counting}). Specifically, the two independent leakage channels $\ket{01} \rightarrow \ket{0r}$ and $\ket{11} \rightarrow \ket{W}$ each contribute two dimensions (corresponding to the real and imaginary parts of the matrix element), while the nonlinear phase contributes another dimension. In this model, $\ket{10}$ is related to $\ket{01}$ by symmetry and does not contribute additional error channels, while $\ket{00}$ does not experience any dynamics. This explains the observed rank of five.

This greatly simplifies the experimental optimization: only the component $\delta\vec{\Omega}_{\parallel}$ needs to be corrected, and the basis vectors of $V$ define a set of orthogonal optimization directions (Fig.~\ref{fig:Fig1}g). In Fig.~\ref{fig:Fig1}h, we perform simulations comparing this approach with three other optimization schemes used in the literature, and consider two aspects: convergence rate and error floor. Our approach of iteratively optimizing Hessian eigenvectors gives rapid convergence with no fidelity floor. Directly optimizing the polynomial coefficients $\alpha_k$ in a chosen basis~\cite{ma_high-fidelity_2023} can also reach the optimum, provided that the basis set is sufficiently large. However, this approach converges much more slowly because the scanned coefficients are generally coupled to one another. In contrast, optimizing the parameters of an analytic ansatz can lead to very slow convergence and an effective error floor if the parameters do not have a large projection onto all axes of the principal space, which is the case for the ansatz in Refs.~\cite{evered_high-fidelity_2023, muniz_high-fidelity_2025}. Optimizing the principal components within this parameterization can improve the convergence~\cite{muniz_high-fidelity_2025}, but the effective error floor remains.
The magnitude of the error floor depends on the magnitude of the initial waveform error that is orthogonal to the control vectors generated by the analytic pulse parameters (Fig.~\ref{fig:Fig1}i).

Connecting the dimension of the principal space to physical error channels also clarifies which Hamiltonian errors can be corrected by the same scan. If a small Hamiltonian perturbation only changes the amplitudes of the leakage, mixing, or phase-error channels already represented in the Hessian eigenvectors, then its leading-order effect lies within the same low-dimensional correctable space and can be removed by scanning along those directions (Appendix~\ref{sec:hamiltonian_error}).

\section{Experimental implementation}
\label{sec:hessian_benchmark}

\begin{figure}
    \centering
    \includegraphics{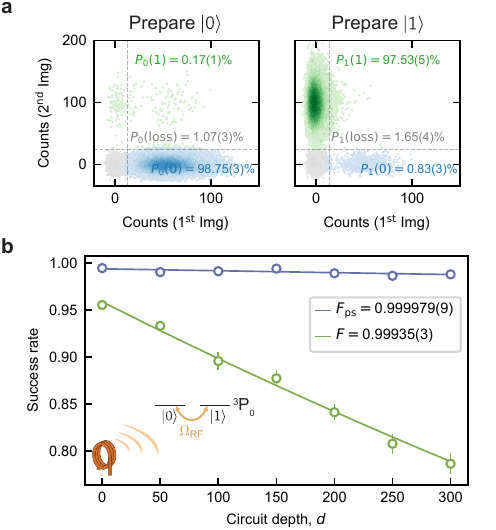}
    \caption{
    \textbf{Gate benchmarking with three-outcome measurement.}
    (a) Photon-count distributions collected from two successive images used to measure the atoms in $\ket{0}$ and $\ket{1}$, respectively. Dashed lines indicate the image-classification thresholds, and colors indicate the assigned outcomes $\ket{0}$, $\ket{1}$, and loss.
    (b) Randomized benchmarking of RF single-qubit gates. The raw fidelity is $F = 0.99935(3)$ (green), which increases to $F_{\rm ps} = 0.999979(9)$ (blue) after postselecting on no detected atom loss.
    In all panels, error bars indicate $1\sigma$ uncertainties.
    }
    \label{fig:SSR_SQRB}
\end{figure}

\begin{figure*}
    \centering
    \includegraphics{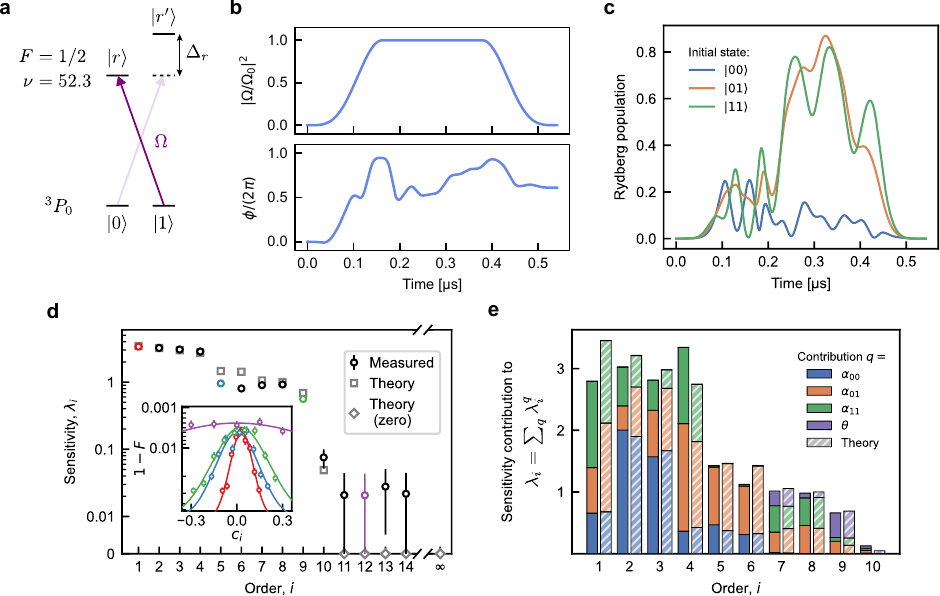}
    \caption{
    \textbf{Verification of the low-rank nature of the fidelity Hessian.}
    (a) Four-level model of the states involved in the Rydberg gate. 
    (b) Numerically optimized gate waveforms for the amplitude-robust gate.
    (c) Rydberg population during the gate when starting from $\ket{00}$, $\ket{01}$, and $\ket{11}$. The small Zeeman splitting between $\ket{r}$ and $\ket{r'}$ results in a significant Rydberg population when starting in $\ket{00}$.
    (d) Comparison between measured (circles) and calculated (squares) sensitivities $\lambda_i$ of the CZ gate fidelity to distortions along the Hessian eigenvectors. Inset: representative measurements used to extract the sensitivity, with colors matching the highlighted points in the main figure. The fidelity is extracted using a randomized benchmarking sequence following Ref.~\cite{peper_spectroscopy_2025}. Error bars indicate $1\sigma$ uncertainties.
    (e) Decomposition of the measured and calculated sensitivities into contributions from leakage errors from different computational states $\lambda_i^{\alpha_{00}}$, $\lambda_i^{\alpha_{01}}$, and $\lambda_i^{\alpha_{11}}$, and coherent error $\lambda_i^{\theta}$. Solid bars denote experimental values (see Appendix~\ref{sec:rank_counting} for measurement details), while hatched bars denote the corresponding theoretical predictions.
    }
    \label{fig:Fig3}
\end{figure*}

\begin{figure*}[t!]
    \centering
    \includegraphics[width=6.75in]{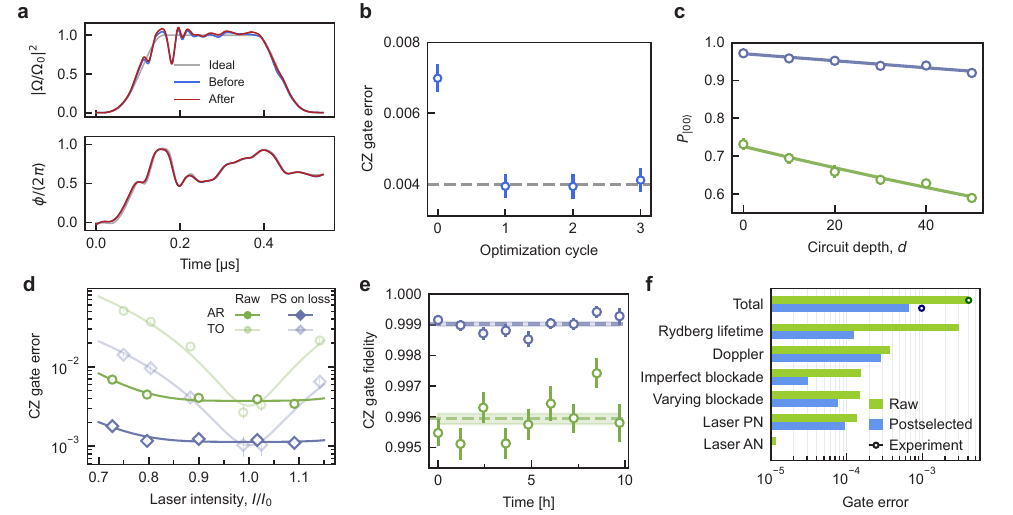}
    \caption{\textbf{Optimization and benchmarking of amplitude-robust CZ gates.} 
    (a) Ideal gate waveform, and measured gate waveforms before and after optimization. The measured waveforms are both significantly distorted by the acousto-optic modulator.
    (b) CZ gate error after several optimization cycles, each consisting of one scan of all 10 Hessian eigenvector coefficients. The dashed line represents the simulated gate error.
    (c) Echoed RB characterization of the AR CZ gate. The gate error is $\varepsilon = 4.0(5)\times10^{-3}$ (green) before loss postselection, which is reduced to $\varepsilon_{\rm ps} = 1.0(2)\times10^{-3}$ (blue) after loss postselection.
    (d) Sensitivity of the AR CZ gate and non-robust time-optimal (TO) CZ gate to changes in the laser intensity, $I/I_0$. The curves show phenomenological fits with quadratic scaling in $\Delta I/I_0$ for the TO gate and quartic scaling for the AR gate, where $\Delta I \equiv I-I_0$.
    (e) Long-term stability of the optimized AR CZ gate. 
    (f) Simulated error budget for the AR CZ gate. ``Laser PN" denotes laser phase noise, ``Laser AN" denotes laser intensity noise.
    In all panels, error bars indicate $1\sigma$ uncertainties.
}
    \label{fig:fidelity}
\end{figure*}

We now turn to the experimental implementation of our optimization approach. We briefly summarize the experimental context. We study qubits encoded in the nuclear spin sublevels of the $^3P_0$ metastable-state manifold of $^{171}$Yb, with $\ket{0}\equiv \ket{^3P_0,m_F=-1/2}$ and $\ket{1}\equiv \ket{^3P_0,m_F=1/2}$~\cite{ma_high-fidelity_2023, lis_midcircuit_2023}. One advantage of this encoding is that gate errors are biased towards transitions outside of the computational subspace, which can be beneficial for error correction whether detected directly~\cite{wu_erasure_2022, sahay2023high, zhang_leveraging_2025} or later in the form of qubit loss~\cite{perrin_quantum_2025, baranes_leveraging_2026, yu2026taming, bluvstein_fault-tolerant_2026}. Here, we characterize gates using a three-outcome measurement that distinguishes $\ket{0}$, $\ket{1}$, and loss~\cite{lis_midcircuit_2023, li_fast_2025, norcia2023midcircuit, chow2024circuit, bluvstein_fault-tolerant_2026}. 
Using a measurement scheme similar to Ref.~\cite{li_fast_2025}, we acquire two successive images (respectively named $1^{\text{st}}$ and $2^{\text{nd}}$ in Fig.~\ref{fig:SSR_SQRB}a) to separately determine the $\ket{0}$ and $\ket{1}$ population. We observe the correct state with probability $98.14(3)\%$, which rises to $99.49(2)\%$ after discarding loss events.
As a demonstration, we apply this measurement to single-qubit randomized benchmarking. The extracted single-qubit gate error is $6.5(3)\times10^{-4}$, while the error postselected on the absence of loss is reduced to $2.1(9)\times10^{-5}$. This indicates that over 95\% of the single-qubit gate error comes from leakage.

For the two-qubit gate, we use an amplitude-robust (AR) controlled-$Z$ gate designed by optimal control. The AR design is robust to spatial intensity inhomogeneity and slow temporal drifts of the laser power~\cite{jandura_optimizing_2023, fromonteil_protocols_2023, zhang_leveraging_2025}, making it attractive for achieving uniform gate fidelities over large gate zones with constrained laser power. In addition to enabling this robustness, optimal control also allows us to design a gate that tolerates other nearby Rydberg states.

This gate is implemented through the $6s\nu s$, $\nu=52.3$, $F=1/2$ Rydberg manifold~\cite{peper_spectroscopy_2025}. In our apparatus, the Rydberg-beam polarization and achievable Zeeman splitting are constrained by optical access and available magnetic-field strength. The Rydberg beam is linearly polarized perpendicular to the magnetic field, and therefore decomposes into equal $\sigma^-$ and $\sigma^+$ components: the $\sigma^-$ component resonantly drives $\ket{1}\rightarrow\ket{r}\equiv\ket{\nu=52.3,m_F=-1/2}$, while the equally strong $\sigma^+$ component also couples $\ket{0}\rightarrow\ket{r'}\equiv\ket{\nu=52.3,m_F=+1/2}$ (Fig.~\ref{fig:Fig3}a). The experimentally available magnetic-field strength in our apparatus limits the Zeeman splitting between these two Rydberg states to $\Delta_r=2\pi\times16.1$~MHz, only a few times larger than the gate Rabi frequency $\Omega_0=2\pi\times6.0$~MHz (which corresponds to approximately 60~mW of 302~nm laser power for a beam waist on the atoms of $w_0 = 12~\mu$m defined at $1/e^2$), such that the dynamics of the $\ket{00}$ state can no longer be neglected.
The resulting gate design and state populations during the gate are shown in Fig.~\ref{fig:Fig3}b and c.

The more complicated level structure introduces additional leakage error channels, and the Hessian rank increases to 10 (Appendix~\ref{sec:rank_counting}). To validate the low-rank nature of the fidelity Hessian, we measure the gate sensitivity along the 10 principal directions and four randomly chosen directions in the null space. The experimental sensitivities track the predicted sensitivities, with a clear distinction between the measured sensitivity of the lowest principal eigenvector and those of the null eigenvectors (Fig.~\ref{fig:Fig3}d). In separate benchmarking experiments, we can measure the type of error caused by perturbing each eigenmode, finding good agreement with the theory prediction (Fig.~\ref{fig:Fig3}e). We note that most eigenmodes are dominated by leakage, suggesting that future calibration routines may be simplified by monitoring selected error channels for these modes.

Next, we optimize the performance of the two-qubit gate by iteratively scanning the coefficient along each of the 10 principal Hessian eigenvectors. We benchmark the gate fidelity using an echoed global randomized benchmarking sequence~\cite{evered_high-fidelity_2023}, which cancels sensitivity to single-qubit phase errors and preserves the number of single-qubit operations. 
Before optimization, the waveform is significantly distorted by the finite bandwidth of the acousto-optic modulator (Fig.~\ref{fig:fidelity}a), and the gate error rate is $\varepsilon=7.0(4)\times10^{-3}$, roughly double the predicted value. The optimization protocol converges to the expected gate error of $\varepsilon_{\mathrm{raw}}=4.0(5)\times10^{-3}$ after a single iteration (Fig.~\ref{fig:fidelity}b). Postselecting on the absence of loss reduces the error to $\varepsilon_{\mathrm{ps}}=1.0(2)\times10^{-3}$ (Fig.~\ref{fig:fidelity}c), corresponding to an erasure fraction of $0.75(6)$. Importantly, the optimized waveform is only slightly modified from the initial waveform, demonstrating that our optimization approach isolates the low-dimensional waveform space relevant for the gate fidelity.

A key advantage of the AR gate is that its performance does not rely on precise control of the laser intensity, and we directly test this robustness by deliberately varying the laser power. For comparison, we also study the non-robust time-optimal gate, optimized using the same technique, which achieves a nominal gate error of $\varepsilon_{\mathrm{raw}}=2.7(5)\times10^{-3}$ before loss postselection, and gate error of $\varepsilon_{\mathrm{ps}}=1.0(3)\times10^{-3}$ after loss postselection~\cite{jandura_time-optimal_2022}. The time-optimal gate rapidly degrades away from its calibrated intensity, while the AR gate fidelity is essentially unchanged for laser power variations up to 20\%~(Fig.~\ref{fig:fidelity}d). Benefiting from this robustness, the AR gate performance remains stable over a $10$-hour period without waveform reoptimization, achieving an average error of $\overline{\varepsilon}_{\mathrm{raw}}=4.1(2)\times10^{-3}$ before loss postselection, and $\overline{\varepsilon}_{\mathrm{ps}}=9.8(7)\times10^{-4}$ after loss postselection (Fig.~\ref{fig:fidelity}e). The optimized gate performance is consistent with a detailed numerical model based on independently measured parameters, indicating that the dominant sources of error are Rydberg decay and Doppler shifts (Fig.~\ref{fig:fidelity}f). We attribute the small discrepancy in the postselected error probability to the difficulty of quantitatively modeling the full Rydberg interaction, including unwanted excitation to $\ket{r'}$ (Appendix~\ref{sec:error_model}). 

\begin{figure}
    \centering
    \includegraphics{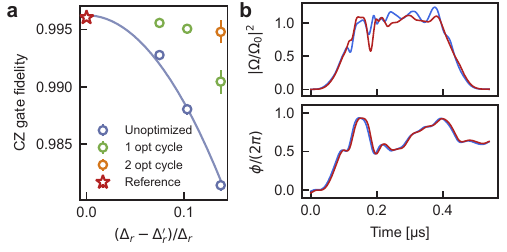}
    \caption{
    \textbf{Optimizing gates with Hamiltonian parameter errors.}
    (a) CZ gate fidelity versus reduction of the Rydberg Zeeman splitting before and after Hessian optimization. A single optimization cycle along the original Hessian eigenvectors eliminates most of the added error, with an additional cycle needed at the largest miscalibration.
    (b) Optimized laser intensity and phase profiles for the nominal case (red) and the reduced-field case (blue, $1-\Delta_r'/\Delta_r=0.14$).
    Error bars indicate $1\sigma$ uncertainties.
    }
    \label{fig:h_error}
\end{figure}

Finally, we demonstrate that many parameter errors in the Hamiltonian model used to design the gate can also be corrected using closed-loop feedback from the experiment within the same low-rank subspace. As an example, we consider the Zeeman splitting to the unwanted $\ket{r'}$ Rydberg level, $\Delta_r$. The gate is designed for $\Delta_r = 2.69\,\Omega_0$, and is quite sensitive to variations in this parameter as a non-trivial amount of population is excited to $\ket{r'}$ from $\ket{0}$ during the gate (as shown in Fig.~\ref{fig:Fig3}c). To mimic a calibration error in this parameter, we deliberately lower the magnetic field while adjusting the laser frequency to match the $\ket{1}\leftrightarrow\ket{r}$ transition. While the gate error is substantially increased by the parameter change, applying the same optimization protocol with the original Hessian eigenvectors recovers most of the lost fidelity (Fig.~\ref{fig:h_error}a). The optimized waveform is significantly different from the nominal waveform (Fig.~\ref{fig:h_error}b). In the case of the largest deviation, the optimization requires more than one round to converge. This multi-round behavior is consistent with numerical calculations: at this level of miscalibration, the principal space has shifted appreciably from that computed at the nominal field, pushing the correction somewhat beyond the strictly perturbative regime. The nominal Hessian directions therefore remain useful sensitive directions, but are no longer optimally aligned with the local error response. A detailed theory of which types of errors can be corrected this way is presented in Appendix~\ref{sec:hamiltonian_error}.

\section{Discussion and Conclusion}

The low-rank Hessian optimization approach presented here is based on leading-order perturbation theory, and its effectiveness therefore requires that the initial distortion is not too large. In simulations, we find that the convergence region is reasonably large: using random initial distortions with different amplitudes and shapes, the method can reliably recover gate errors below $10^{-5}$ after several cycles even when the initial error is as high as $\sim 10^{-1}$.

A complementary approach to calibrating optimal control gates is to directly measure the pulse errors and pre-compensate them. This has been demonstrated for neutral atoms~\cite{ma_high-fidelity_2023} and is extensively used for superconducting qubits~\cite{rol_time-domain_2020, hellings_calibrating_2025}. This approach has the benefit of not relying on the qubits for feedback, but is subject to measurement and modeling errors. In the present work, we do not apply any pre-compensation; instead, we start from the initially implemented waveform and directly optimize based on the measured gate fidelity. We nevertheless expect the two methods to be complementary: direct waveform feedback can bring a strongly distorted pulse into the perturbative regime, after which low-rank Hessian optimization can efficiently correct the remaining errors.

We now discuss how to further improve the gate. In the present geometry, unwanted coupling to the additional Rydberg state $\ket{r'}$ complicates the dynamics, adds error channels, and increases the number of calibration directions. A larger magnetic field together with a more favorable laser-polarization geometry would allow a nearly pure $\sigma^-$ drive, suppressing this extra coupling and bringing the gate closer to the ideal three-level picture. In this regime, our model predicts that a time-optimal gate driven at $\Omega=2\pi\times14~\mathrm{MHz}$, together with modest improvements in atomic temperature and laser-intensity noise, can reach fidelities above $0.999$, with more than $90\%$ of the remaining error appearing as leakage. This requires increasing the laser intensity by a factor of 2.7.

Finally, we emphasize that the low-rank Hessian theory and optimization protocol should be broadly applicable to other gates and other types of qubits. It may be particularly applicable to solid-state qubits, which suffer from large control signal distortion from long cryogenic signal chains and qubit-to-qubit variation in Hamiltonian parameters from device fabrication uncertainty.

\section{Acknowledgments}
We acknowledge Yiyi Li, Michael Peper, Yicheng Bao, Pranav Mathur, Matteo Bergonzoni and Guido Pupillo for helpful conversations. This work was supported by the Army Research Office (W911NF-24-10358), DARPA MeasQuIT (HR00112490363), the Office of Naval Research (N00014-23-1-2621, N00014-26-1-2102), and the National Science Foundation through the CAREER program (PHY-2047620) and the Center for Robust Quantum Simulation (OMA-2120757).

\emph{Note:} While completing this work, we became aware of complementary work on high-fidelity neutral atom gates~\cite{evered2026high}.

\appendix
\setcounter{figure}{0}
\renewcommand{\thefigure}{A\arabic{figure}}
\renewcommand{\theHfigure}{appendix.\arabic{figure}}

\input{appendix/experimental_methods}
\input{appendix/general_hessian_theory}
\input{appendix/2q_numerical_model}
\input{appendix/three_outcome_measurement}
\input{appendix/rb_sequence}

\bibliography{ref}

\end{document}

%% file: appendix/experimental_methods.tex
\section{Experimental methods} \label{sec:Exp_methods}
\begin{figure}[b]
    \centering
    \includegraphics{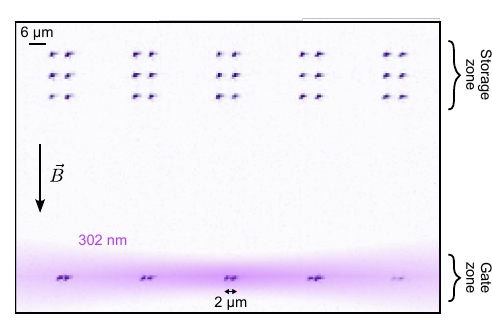}
    \caption{
    \textbf{Average fluorescence image of the atomic array.}
    The array is divided into a storage zone and a gate zone (see text). The quantization axis $\vec{B}$ is perpendicular to the $302~\mathrm{nm}$ beam.
    }
    \label{fig:fluo_atom_image}
\end{figure}

We generate an array of 40 spatial-light-modulator (SLM)-defined tweezers that are loaded from a three-dimensional magneto-optical trap (MOT) operating on the $^1S_0 \rightarrow {}^3P_1$ transition. The SLM tweezer array is divided into a storage zone and a gate zone, with the latter consisting of 10 traps illuminated by a tightly focused $302~$nm UV gate laser (see Fig.~\ref{fig:fluo_atom_image}). A crossed acousto-optic deflector (AOD) tweezer array is used to dynamically move atoms between the SLM traps. This enables the preparation of defect-free arrays and the transport of atoms to and from the gate zone during an experimental sequence. Details of the experimental chamber~\cite{Saskin.2019}, the SLM tweezer array, and the AOD-controlled moving tweezer array~\cite{zhang_leveraging_2025} are described in previous work. The qubits are initialized in the two nuclear-spin states of the metastable $6s6p\,{}^3P_0$ manifold by optically pumping atoms from the ground state, as described in Ref.~\cite{ma_high-fidelity_2023}. Qubit readout is performed using a three-outcome measurement, which is adapted from Ref.~\cite{li_fast_2025} and discussed in more detail in Appendix~\ref{sec:Three-outcome}.

\begin{figure}
    \centering
    \includegraphics{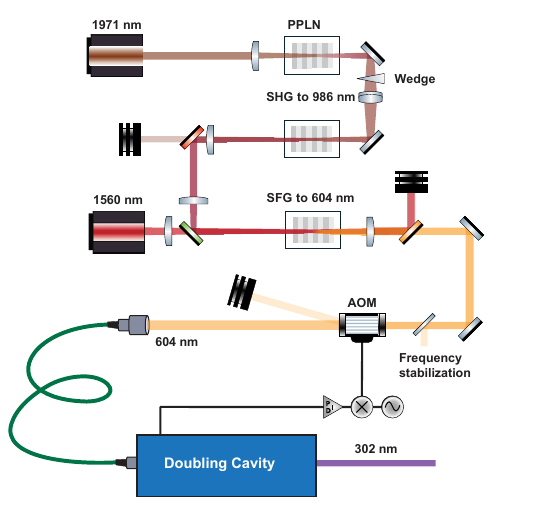}
    \caption{
    \textbf{Optical setup for generating 302~nm light.}
    The 302~nm light used for Rydberg excitation is generated from amplified 1971~nm and 1560~nm fiber-laser sources through SHG, SFG, and cavity-enhanced frequency doubling.
    }
    \label{fig:uv_setup}
\end{figure}

In this work, we begin each experimental sequence by preparing a set of five dimers, which are then moved to the gate zone for the single- and two-qubit gates presented in the main text. In the gate zone, the two atoms within each dimer are separated by 2.0~$\mu$m, while neighboring dimers are separated by 25~$\mu$m. The $\mathrm{CZ}$ gates are driven by a tightly focused 302~nm laser, with a $1/e^2$ radius $w_0 = 12~\mu$m, that couples the $\ket{1}$ state to the Rydberg state $\ket{r} \equiv \ket{\nu = 52.3, F=1/2, m_F = -1/2}$. The UV light is generated by nonlinear frequency conversion of amplified fiber-laser sources (Fig.~\ref{fig:uv_setup}). We first frequency-double a 1971 nm laser (Precilasers) to 986 nm using two periodically poled lithium niobate (PPLN) second-harmonic generation (SHG) crystals in series to increase the available 986 nm power. This light is then combined with 1560 nm light (Precilasers) in a sum-frequency generation stage, producing 604 nm light. Finally, the 604 nm light is frequency doubled in an enhancement cavity (Toptica), generating the 302 nm light used for Rydberg excitation.

%% file: appendix/general_hessian_theory.tex
\section{The general theory of low-rank fidelity Hessians} 
\label{sec:general_hessian_theory}

In this section, we introduce the general theory underlying the low-rank Hessian structure of gate fidelity. Consider a quantum gate generated by an ideal Hamiltonian $H_0(t)$ over the time interval $t\in[0,T]$. This Hamiltonian gives an ideal unitary evolution $U_0(t)$, with $U_0(T)$ implementing the desired gate. We now suppose that control distortions perturb the system through a set of control channels, so that the Hamiltonian becomes
\begin{equation} \label{eq:ham_general}
    H(t)=H_0(t)+\sum_{\mu} s_\mu(t)O_\mu(t),
\end{equation}
where $O_\mu(t)$ are operators describing how each distortion channel couples to the system, and the real functions $s_\mu(t)$ specify the corresponding distortion waveforms. The perturbed Hamiltonian generates the actual evolution $U(t)$.

Then the interaction-picture evolution operator,
\begin{equation} \label{eq:interation_unitary}
    U_I(t)=U_0^\dagger(t)U(t),
\end{equation}
removes the ideal evolution and describes the error accumulated from the control distortions. Taking $\hbar=1$, and defining
\begin{equation}
    O_{\mu,I}(t)=U_0^\dagger(t)O_\mu(t)U_0(t),
\end{equation}
the Dyson expansion can be written compactly as
\begin{equation} \label{eq:dyson_expansion}
    U_I(t)=\mathbb{1}-iK_1(t)-K_2(t)+\mathcal{O}(s^3),
\end{equation}
where
\begin{widetext}
\begin{equation} \label{eq:k_definition}
    K_1(t)=\int_0^t \mathrm{d}t_1
    \sum_{\mu} s_\mu(t_1)O_{\mu,I}(t_1), \qquad
    K_2(t)=\int_0^t \mathrm{d}t_1
    \int_0^{t_1}\mathrm{d}t_2
    \sum_{\mu,\nu}
    s_\mu(t_1)s_\nu(t_2)
    O_{\mu,I}(t_1)O_{\nu,I}(t_2).
\end{equation}
\end{widetext}

The gate fidelity is determined by the final interaction-picture evolution $U_I(T)$. Since the ideal gate has been removed in this frame, a perfect implementation corresponds to $U_I(T)=\mathbb{1}$ within the computational subspace. In general, the full Hilbert space contains both the $d$-dimensional computational subspace and additional noncomputational states. Let $P$ be the projector onto the computational subspace, and let $Q=\mathbb{1}-P$ project onto the leakage space. The average gate fidelity is
\begin{equation} \label{eq:f_vs_unitary}
    \mathcal{F}
    =
    \frac{
    \left|\Tr[PU_I(T)P]\right|^2
    +
    \Tr[PU_I(T)PU_I^\dagger(T)]
    }
    {d(d+1)} .
\end{equation}
Because $U_I(T)$ is generated by the distortion waveforms $s_\mu(t)$, the gate fidelity is a functional of these waveforms, which we denote by $\mathcal{F}[\{s_\mu\}]$. Substituting the Dyson expansion of Eqs.~\eqref{eq:dyson_expansion} and~\eqref{eq:k_definition}, and keeping terms up to second order gives
\begin{widetext}
\begin{align}
    \mathcal{F}[\{s_\mu\}]
    &=
    1
    -
    \frac{1}{2}
    \sum_{\mu,\nu}
    \int_0^T \! \dd t_1
    \int_0^T \! \dd t_2\,
    s_\mu(t_1)
    \mathcal{H}_{\mu\nu}(t_1,t_2)
    s_\nu(t_2)
    +\mathcal{O}(s^3), \label{eq:f_vs_hessian}
    \\
    \mathcal{H}_{\mu\nu}(t_1,t_2)
    &=
    \frac{2}{d+1}
    \Tr[
    \tilde{O}_{\mu}(t_1)
    \tilde{O}_{\nu}(t_2)
    ]
    +
    \frac{2}{d}
    \Tr[
    L_\mu(t_1)
    L_\nu^\dagger(t_2)
    ] .  \label{eq:hessian_form}
\end{align}
\end{widetext}
The Hessian kernel $\mathcal{H}_{\mu\nu}(t_1,t_2)$ is written in terms of
\begin{align}
    \tilde{O}_\mu(t)
    &=
    PO_{\mu,I}(t)P
    -
    \frac{\Tr[PO_{\mu,I}(t)P]}{d}P, \label{eq:traceles_o} \\
    L_\mu(t)
    &=
    PO_{\mu,I}(t)Q . \label{eq:leakage_l}
\end{align}
Here $\tilde{O}_\mu(t)$ is the traceless part of $O_{\mu,I}(t)$ within the computational subspace, while $L_\mu(t)$ describes the coupling between the leakage and computational subspaces.

To interpret Eq.~\eqref{eq:f_vs_hessian}, it is useful to view it as the continuous-waveform analogue of an ordinary finite-dimensional quadratic form. The collection of distortion waveforms $\{s_\mu(t)\}$ plays the role of a vector $\vec{s}$. If there are $N_c$ control channels, then $\vec{s}$ lives in the direct sum of $N_c$ waveform spaces, one for each value of $\mu$. The pair $(\mu,t)$ is therefore analogous to a single vector index. Similarly, $\mathcal{H}_{\mu\nu}(t_1,t_2)$ plays the role of a matrix element, with $(\mu,t_1)$ and $(\nu,t_2)$ labeling its two indices. In this notation, Eq.~\eqref{eq:f_vs_hessian} has the same structure as
\begin{equation}
    \mathcal{F}(\vec{s})
    =
    1-\frac{1}{2}\vec{s}^{\,T}\mathcal{H}\vec{s}
    +\mathcal{O}(s^3),
\end{equation}
except that the sums over vector indices are replaced by sums over control channels and integrals over time. Thus, the basic structure is the familiar quadratic form, but now in an infinite-dimensional waveform space.

This analogy also allows us to define Hessian eigendirections in the same way as for an ordinary matrix. The Hessian kernel can be diagonalized through the integral eigenvalue problem
\begin{equation} \label{eq:eigen_problem}
    \sum_\nu \int_0^T \! \dd t_2\,
    \mathcal{H}_{\mu\nu}(t_1,t_2)
    v_{i,\nu}(t_2)
    =
    \lambda_i v_{i,\mu}(t_1).
\end{equation}
The eigenfunction $\vec{v}_i=\{v_{i,\mu}(t)\}$ is therefore an eigendirection in the waveform-distortion space, while the eigenvalue $\lambda_i$ gives the second-order sensitivity of the fidelity along that direction. The rank of the Hessian kernel is defined analogously as the number of nonzero eigenvalues, or equivalently the number of independent eigendirections with nonzero sensitivity.

To see why this Hessian has finite rank, choose an orthonormal basis
$\{\ket{n}\}_{n=1}^D$ for the full Hilbert space, with
$\ket{1},\ldots,\ket{d}$ spanning the computational subspace and
$\ket{d+1},\ldots,\ket{D}$ spanning the leakage subspace. Expanding the traces in Eq.~\eqref{eq:hessian_form} gives
\begin{widetext}
\begin{align}
    \Tr[
    \tilde{O}_{\mu}(t_1)
    \tilde{O}_{\nu}(t_2)
    ]
    &=
    \sum_{m,n=1}^{d}
    \bra{m}\tilde{O}_{\mu}(t_1)\ket{n}
    \bra{n}\tilde{O}_{\nu}(t_2)\ket{m}
    =
    \sum_{m,n=1}^{d}
    \chi^{\mathrm{coh}}_{mn,\mu}(t_1)
    \chi^{\mathrm{coh}*}_{mn,\nu}(t_2),
    \\
    \Tr[
    L_\mu(t_1)
    L_\nu^\dagger(t_2)
    ]
    &=
    \sum_{m=1}^{d}
    \sum_{\ell=d+1}^{D}
    \bra{m}O_{\mu,I}(t_1)\ket{\ell}
    \bra{\ell}O_{\nu,I}(t_2)\ket{m}
    =
    \sum_{m=1}^{d}
    \sum_{\ell=d+1}^{D}
    \chi^{\mathrm{leak}}_{m\ell,\mu}(t_1)
    \chi^{\mathrm{leak}*}_{m\ell,\nu}(t_2).
\end{align}
\end{widetext}
Here we have defined the waveform-space channel vectors
$\chi^{\mathrm{coh}}_{mn,\mu}(t)=\bra{m}\tilde{O}_{\mu}(t)\ket{n}$ and
$\chi^{\mathrm{leak}}_{m\ell,\mu}(t)=\bra{m}O_{\mu,I}(t)\ket{\ell}$, whose components are labeled by the control channel $\mu$ and time $t$. Substituting these expressions into Eq.~\eqref{eq:hessian_form}, the Hessian kernel is a finite sum of outer products of these vectors, or schematically
\begin{equation} \label{eq:hessian_vs_chi}
    \mathcal{H}
    =
    \frac{2}{d+1}
    \sum_{m,n=1}^{d}
    \vec{\chi}^{\mathrm{coh}}_{mn}
    \vec{\chi}^{\mathrm{coh}\,\dagger}_{mn}
    +
    \frac{2}{d}
    \sum_{m=1}^{d}
    \sum_{\ell=d+1}^{D}
    \vec{\chi}^{\mathrm{leak}}_{m\ell}
    \vec{\chi}^{\mathrm{leak}\,\dagger}_{m\ell}.
\end{equation}
This outer-product form makes the finite-rank structure straightforward. For any input waveform $\vec{s}$, the output $\mathcal{H}\vec{s}$ lies in the span of the channel vectors $\vec{\chi}^{\mathrm{coh}}_{mn}$ and $\vec{\chi}^{\mathrm{leak}}_{m\ell}$. Conversely, any waveform orthogonal to this span has zero overlap with every outer-product term and therefore lies in the null space of the Hessian. The Hessian rank is therefore bounded by the number of independent real directions contained in these channel vectors.

To obtain an explicit upper bound, we now count the number of independent real directions. The diagonal coherent vectors $\chi_{mm,\mu}^{\mathrm{coh}}(t)$ are real by Hermiticity of $\tilde{O}_\mu(t)$ and correspond to phase errors on the computational states. Because $\tilde{O}_\mu(t)$ is traceless within the computational subspace [Eq.~\eqref{eq:traceles_o}],
\begin{equation}
    \sum_{m=1}^{d} \chi_{mm,\mu}^{\mathrm{coh}}(t)=0 .
\end{equation}
Thus only $d-1$ phase-error channels are independent; the missing direction is the global phase, which does not affect the gate fidelity.

The off-diagonal coherent vectors $\chi_{mn,\mu}^{\mathrm{coh}}(t)$ describe mixing between computational states. Since $\chi_{nm}^{\mathrm{coh}}=\chi_{mn}^{\mathrm{coh}*}$, each pair $m<n$ gives one complex channel, or two real directions, contributing at most $d(d-1)$ to the Hessian rank. The leakage vectors $\chi_{m\ell,\mu}^{\mathrm{leak}}(t)$ are also complex; there are $d(D-d)$ such channels, contributing at most $2d(D-d)$ real directions. 

Combining these contributions, the Hessian rank is bounded by
\begin{equation} \label{eq:2dD-d2-1}
    \rank(\mathcal{H})\leq
    (d-1)+d(d-1)+2d(D-d)
    =
    2dD-d^2-1 .
\end{equation}
The principal space $V$ defined in the main text is therefore the real span of these channel vectors,
\begin{equation}
    V =
    \mathrm{span}_{\mathbb{R}}\,
    \{
    \mathrm{Re}[\vec{\chi}_{mn}^\mathrm{coh}],
    \mathrm{Im}[\vec{\chi}_{mn}^\mathrm{coh}],
    \mathrm{Re}[\vec{\chi}_{m\ell}^\mathrm{leak}],
    \mathrm{Im}[\vec{\chi}_{m\ell}^\mathrm{leak}]
    \}.
\end{equation}
If the full Hilbert space is the computational space, $D=d$, this reduces to the general bound $\rank(\mathcal{H})\leq D^2-1$ in the main text and Refs.~\cite{ho_landscape_2009, berger_dimensionality_2024}. This unitary-fidelity result should be distinguished from previous Hessian-rank bounds for state-transfer fidelities, where the objective is sensitive only to errors in a single target state and the Hessian rank is correspondingly much smaller, bounded by $2(D-1)$~\cite{rabitz_topology_2006, shen_quantum_2006}.

The above derivation also gives a useful physical interpretation of how waveform distortions affect the gate. The channel vectors $\vec{\chi}^{\mathrm{coh}}_{mn}$ and $\vec{\chi}^{\mathrm{leak}}_{m\ell}$ are essentially the waveform kernels that determine the corresponding matrix elements of the first-order error operator $K_1(T)$ in Eq.~\eqref{eq:dyson_expansion}. Schematically, the relevant part of the interaction-picture evolution can be written as
\begin{equation}
    U_I(T)
    =
    \mathbb{1}
    -i\sum_{m,n}
    \left(\vec{s}^{\,T}\vec{\chi}_{mn}\right)
    \ket{m}\bra{n}
    +\mathcal{O}(s^2),
\end{equation}
where $\vec{\chi}_{mn}$ denotes the waveform-space vector associated with the matrix element $\chi_{mn,\mu}(t)=\bra{m}O_{\mu,I}(t)\ket{n}$. It can be shown that, to leading order, the correction to the fidelity is determined entirely by the relevant matrix elements of $K_1(T)$: diagonal computational-subspace elements give phase-error channels, off-diagonal computational-subspace elements give mixing channels, and computational--leakage elements give leakage channels. Evolution entirely within the leakage subspace does not contribute at this order because the system starts in the computational subspace. This formalism will also be useful for understanding which Hamiltonian errors can be corrected by waveform optimization, as illustrated more explicitly in Appendix~\ref{sec:hamiltonian_error}.

\section{Rank counting and fidelity decomposition}
\label{sec:rank_counting}

Eq.~\eqref{eq:2dD-d2-1} gives a general finite bound on the rank of the fidelity Hessian. In specific systems, however, this bound can be loose because symmetries and selection rules eliminate many error channels. A tighter bound can often be obtained by identifying the independent physical error channels and applying the same rank-counting logic:
\begin{equation}
    \rank(\mathcal{H})
    \leq
    N_{\mathrm{phase}}
    +
    2(N_{\mathrm{mixing}} + N_{\mathrm{leakage}}),
\end{equation}
where $N_{\mathrm{phase}}$ counts independent real phase-error channels, while $N_{\mathrm{mixing}}$ and $N_{\mathrm{leakage}}$ count independent complex mixing and leakage channels, respectively.

For example, in the three-level CZ model shown in Fig.~\ref{fig:Fig1}a and b, there is no mixing between different computational states. The only independent leakage channels are
\begin{align*}
    \ket{01} &\leftrightarrow \ket{0r}, \\
    \ket{11} &\leftrightarrow \ket{W}
    =
    \frac{\ket{1r}+\ket{r1}}{\sqrt{2}},
\end{align*}
with the $\ket{10}\leftrightarrow\ket{r0}$ channel related to the first one by symmetry, while $\ket{00}$ is uncoupled. Each leakage channel contributes at most two real Hessian directions. After removing the freely tunable single-qubit phase, the only remaining phase-error channel is the controlled phase, $\varphi_{11}-2\varphi_{01}$, whose target value is $\pi$. Thus this model has at most five nonzero Hessian directions: four from the two leakage channels and one from the controlled-phase channel.

For the more complicated AR CZ gate shown in Fig.~\ref{fig:Fig3}a and b, there is still no mixing between different computational states. However, the independent leakage channels are now
\begin{equation}
    \ket{00}\leftrightarrow\ket{W'},\,
    \ket{01}\leftrightarrow\ket{0r},\,
    \ket{01}\leftrightarrow\ket{r'1},\,
    \ket{11}\leftrightarrow\ket{W}.
\end{equation}
Here $\ket{W'}=(\ket{0r'}+\ket{r'0})/\sqrt{2}$ is the analogous symmetric leakage state. These four complex leakage channels contribute at most eight real Hessian directions. For the phase channels, the phase of $\ket{00}$ can be taken as an irrelevant global phase. In this particular implementation, we choose to fix the relative phases of both $\ket{01}$ and $\ket{11}$, leaving two independent relative phase channels. Therefore, the total Hessian rank is bounded by $4\times2+2=10$.

\begin{figure*}[htp!]
    \centering
    \includegraphics[width=6.75in]{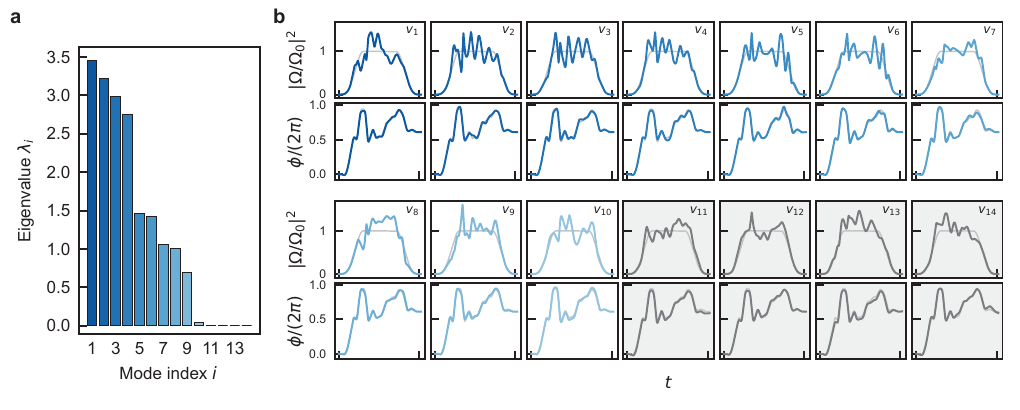}
    \caption{
    \textbf{Eigenspectrum and eigenvectors of the AR gate fidelity Hessian.}
    (a) Computed eigenvalues of the 10 principal Hessian modes, together with 4 representative null-space modes.
    (b) Intensity $|\Omega(t)/\Omega_0|^2$ and phase profile $\phi(t)$ of the AR CZ gate distorted along each eigendirection $\vec{v}_i\equiv\{v_{i,x}(t),v_{i,y}(t)\}$ corresponding to the eigenvalues in (a), with normalization $\int_0^T (v_{i,x}^2+v_{i,y}^2)\,\mathrm{d}t=1$. The distortion strength is $\epsilon=0.6$, with $s_{x/y}(t)=\epsilon v_{i,x/y}(t)$ in Eq.~\eqref{eq:sxsy}. The light gray lines indicate the gate pulse without distortion.
    }
    \label{fig:AR_eigenvecs}
\end{figure*}

We now specialize to the AR CZ gate and show how the abstract channel picture appears in the leakage and phase contributions to the fidelity. In this model, the ideal Hamiltonian $H_0(t)$ has the form
\begin{equation}
    H_0(t)
    =
    \frac{\Omega_0(t)e^{i\phi_0(t)}}{2}\sigma_+
    + \mathrm{h.c.},
\end{equation}
where $\Omega_0(t)$ and $\phi_0(t)$ are the ideal amplitude and phase waveforms. Including the polarization and Clebsch--Gordan coefficients, the raising operator is
\begin{align}
    \sigma_+
    =
    & \ket{0r}\bra{01} - \ket{r'1}\bra{01}
     + \ket{r0}\bra{10} - \ket{1r'}\bra{10} \nonumber\\
    & -\sqrt{2}\ket{W'}\bra{00} + \sqrt{2}\ket{W}\bra{11}.
\end{align}

To describe waveform distortions, we follow Eq.~\eqref{eq:ham_general} and perturb the ideal Hamiltonian in the two drive quadratures:
\begin{equation} \label{eq:sxsy}
    H(t)
    =
    H_0(t)
    +
    s_x(t)O_x(t)
    +
    s_y(t)O_y(t),
\end{equation}
where $s_x(t)$ and $s_y(t)$ describe distortions in the two drive quadratures. In this convention,
\begin{equation}
    O_x(t)
    =
    \frac{\Omega_0(t)}{2}
    \left(\sigma_+ + \sigma_-\right),
    \quad
    O_y(t)
    =
    \frac{i\Omega_0(t)}{2}
    \left(\sigma_+ - \sigma_-\right).
\end{equation}
The factor $\Omega_0(t)$ is included so that the distortion is tapered by the ideal pulse envelope and vanishes when the ideal laser amplitude is zero. With the ideal Hamiltonian and perturbation operators specified, the Hessian kernel follows directly from Eq.~\eqref{eq:hessian_form}. The eigendirections are obtained by solving the integral eigenproblem in Eq.~\eqref{eq:eigen_problem}, which can be done numerically after discretizing the waveform space, for example in a time-bin basis. The eigenspectrum and corresponding eigenvectors are illustrated in Fig.~\ref{fig:AR_eigenvecs}a,b.

To interpret these eigendirections physically, we now rewrite the same leading-order fidelity loss in terms of leakage and phase errors. Since the AR CZ dynamics do not mix different computational states, the ideal evolution within the computational subspace $\{00,01,10,11\}$ is diagonal,
\begin{equation}
    PU_0(T)P=\mathrm{diag}\left(e^{i\varphi_{00}},e^{i\varphi_{01}},e^{i\varphi_{01}},e^{i\varphi_{11}}\right),
\end{equation}
with the CZ condition $\varphi_{11}-2\varphi_{01}+\varphi_{00}=\pi$. Using the interaction-picture error unitary of Eq.~\eqref{eq:interation_unitary}, and removing the common phase of the $\ket{00}$ sector, its computational-subspace projection can be parameterized as
\begin{widetext}
\begin{equation}
    P U_I(T)P=\mathrm{diag}\left(1-\alpha_{00},(1-\alpha_{01})e^{i\theta_{01}},(1-\alpha_{01})e^{i\theta_{01}},(1-\alpha_{11})e^{i\theta_{11}}\right).
\end{equation}
\end{widetext}
Here, $\theta_{01}$ and $\theta_{11}$ are residual phase errors relative to the $\ket{00}$ sector, while $\alpha_{00}$, $\alpha_{01}$, and $\alpha_{11}$ describe the reduction of the return amplitude in each sector. Substituting this form into the average gate fidelity and expanding to leading nonvanishing order gives
\begin{align}
    1-\mathcal{F}&=\frac{1}{2}\alpha_{00}+\alpha_{01}+\frac{1}{2}\alpha_{11}+\epsilon_{\theta}, \label{eq:arcz_f_expansion}\\
    \epsilon_\theta&=\frac{1}{5}\left(\theta_{01}-\frac{\theta_{11}}{2}\right)^2+\frac{1}{10}\theta_{11}^2. \label{eq:arcz_phase_error}
\end{align}
This separates the infidelity into leakage from $\ket{00}$, leakage from the symmetric $\ket{01}$ and $\ket{10}$ sectors, leakage from $\ket{11}$, and residual coherent phase error $\epsilon_\theta$.

We can connect the $\alpha$ and $\theta$ parameters back to the first-order channels $\vec{\chi}$ in $K_1(T)$, introduced in Appendix~\ref{sec:general_hessian_theory}. It can be shown by expanding the sector return amplitude, $(1-\alpha_q)e^{i\theta_q}$, and comparing it with the Dyson-series expansion in Eqs.~\eqref{eq:dyson_expansion} and \eqref{eq:k_definition}, that $\theta$ is determined by the relative diagonal coherent channels,
\begin{equation}
    \theta_q=-\vec{s}^{\,T}\left(\vec{\chi}^{\mathrm{coh}}_{qq}-\vec{\chi}^{\mathrm{coh}}_{00,00}\right),
    \qquad
    q\in\{01,11\},
\end{equation}
while $\alpha$ is determined by the squared first-order leakage amplitudes,
\begin{equation}
    \alpha_q=\frac{1}{2}\sum_{\ell\in L_q}\left|\vec{s}^{\,T}\vec{\chi}^{\mathrm{leak}}_{q\ell}\right|^2,
    \qquad
    q\in\{00,01,11\}.
\end{equation}
For the AR CZ gate, the relevant leakage sets are
\begin{equation}
    L_{00}=\{W'\},\qquad
    L_{01}=\{0r,r'1\},\qquad
    L_{11}=\{W\}.
\end{equation}
The experimentally meaningful quantities $\alpha$ and $\theta$ therefore provide a physical parameterization of the same channel vectors that determine the Hessian in Eq.~\eqref{eq:hessian_vs_chi}. This example also clarifies what we mean by a leakage channel: it is not specified by the initial computational sector alone, but by the transition from that sector to a particular leakage state. Thus the $\ket{01}$ sector contains two independent leakage channels, to $\ket{0r}$ and $\ket{r'1}$. 

Equation~\eqref{eq:arcz_f_expansion} shows that the AR CZ infidelity induced by a waveform distortion separates into three leakage contributions and one coherent phase contribution. Since each contribution is quadratic in the distortion to leading order, the curvature along each normalized Hessian eigendirection can be decomposed in the same way,
\begin{equation}
    \lambda_i
    =
    \lambda_i^{\alpha_{00}}
    +
    \lambda_i^{\alpha_{01}}
    +
    \lambda_i^{\alpha_{11}}
    +
    \lambda_i^{\theta}.
\end{equation}
For each eigendirection, these four terms can be computed by substituting the corresponding distortion mode into the leakage and phase expressions above. This is the decomposition shown in Fig.~\ref{fig:Fig3}e of the main text.

Experimentally, we independently measure the leakage amplitude $\alpha_q$ and phase shift $\theta_q$ while scanning along each Hessian eigendirection. To extract $\alpha_q$ for a leakage channel $q$, we prepare the corresponding initial computational state and apply repeated CZ gates, with each gate followed by autoionization to measure the population leaked out of the computational subspace. To measure $\theta_q$, we prepare a superposition of the relevant computational states and perform Ramsey sequences with multiple inserted CZ gates. We then use Eqs.~\eqref{eq:arcz_f_expansion} and \eqref{eq:arcz_phase_error} to convert the measured $\alpha_q$ and $\theta_q$ into the predicted sensitivity, and compare the result with the directly measured gate sensitivity.

\section{Correction for Hamiltonian errors} 
\label{sec:hamiltonian_error}

The previous Appendices~\ref{sec:general_hessian_theory} and \ref{sec:rank_counting} focused on waveform distortions. The same first-order channel picture also gives a simple criterion for when the Hessian-based scan can correct other Hamiltonian errors.

Consider an implemented Hamiltonian consisting of the ideal control Hamiltonian, a waveform correction, and an additional small Hamiltonian perturbation:
\begin{equation}
    H(t)
    =
    H_0(t)
    +
    \sum_\mu s_\mu(t)O_\mu(t)
    +
    \epsilon H_p(t).
\end{equation}
Here, the correction waveform is restricted to the scanned Hessian eigendirections in the principal space, $\vec{s}=\sum_i c_i\vec{v}_i$. Using the same perturbative expansion as in Eq.~\eqref{eq:dyson_expansion}, the first-order error generator is linear in both the waveform correction and the additional Hamiltonian perturbation:
\begin{equation}
    K_1(T)
    =
    \sum_i c_i K_1^{(i)}(T)
    +
    \epsilon K_1^{(p)}(T).
\end{equation}
Here, $K_1^{(i)}(T)$ is the first-order generator obtained by setting the waveform distortion to $\vec{v}_i$, while $K_1^{(p)}(T)$ is generated by the perturbation $H_p(t)$,
\begin{equation}
    K_1^{(p)}(T)
    =
    \int_0^T
    U_0^\dagger(t) H_p(t) U_0(t)
    \dd t .
\end{equation}
Therefore, the Hamiltonian error can be cancelled to first order using only the scanned eigendirections if, for the relevant channels connected to the computational subspace,
\begin{equation}
    K_1^{(p)}(T)
    \in
    \mathrm{span}_{\mathbb{R}}
    \left\{
    K_1^{(i)}(T)
    \right\}.
\end{equation}
Equivalently, the perturbation must generate only first-order phase, mixing, or leakage channels already accessible through the scanned Hessian directions. If it introduces a new channel outside this span, the restricted optimization cannot remove the leading-order error without adding new control directions.

For the AR CZ gate, the first-order correctable space is spanned by the four complex leakage channels and two real phase channels (Appendix~\ref{sec:rank_counting}). A laser-detuning error provides the simplest correctable example: it is equivalent to an error in the phase chirp of the drive and therefore does not change the channel structure. A less trivial example is an error in the calibrated ratio between the Rydberg-state Zeeman splitting $\Delta_r$ and the UV Rabi frequency $\Omega_0$. For a fractional error in the splitting between $\ket{r}$ and $\ket{r'}$, the perturbation can be written as
\begin{equation}
    \epsilon H_p
    =
    \epsilon\Delta_r
    \left(
    \ket{r'}\bra{r'}\otimes\mathbb{1}
    +
    \mathbb{1}\otimes\ket{r'}\bra{r'}
    \right).
\end{equation}
This perturbation is not itself one of the laser-control quadratures, so its correctability is not guaranteed from the control Hamiltonian alone. However, it does not introduce new first-order error channels; it only changes the weights of the existing phase and leakage channels. It is therefore correctable to leading order by the same Hessian scan, as demonstrated in the main text. In contrast, a perturbation that couples to a new leakage state, or one that introduces mixing between computational states, generally produces new first-order channels and cannot be fully corrected without enlarging the control space.

%% file: appendix/2q_numerical_model.tex
\section{Error model for CZ gates} \label{sec:error_model}

To understand the sources of gate infidelity in our two-qubit gates, we develop a numerical model similar to that used in Refs.~\cite{ma_high-fidelity_2023, peper_spectroscopy_2025}. The simulation is primarily based on the four-level $\{0,1,r,r'\}$ model discussed above, and combines a master-equation treatment with Monte Carlo sampling. The parameters used in the model are determined from independent experiments, while the simulated gate pulse is the designed waveform in Fig.~\ref{fig:Fig3}b. 

We measure the Rydberg-state lifetime to be $T_r=42(2)~\mu\mathrm{s}$ by exploiting the ability to trap Rydberg atoms in ytterbium~\cite{wilson_trapping_2022}. This finite lifetime contributes $\varepsilon_{\mathrm{raw}}=3.1\times10^{-3}$ to the raw gate infidelity. Most of this error, however, appears as leakage out of the qubit subspace: we separately measure that about $90\%$ of the Rydberg-lifetime-induced errors leave the qubit subspace. As a result, after postselection on no detected loss, this contribution is reduced to $\varepsilon_{\mathrm{ps}}=1.2\times 10^{-4}$.

Another important contribution comes from the Doppler effect, which contributes $\varepsilon_{\mathrm{raw}}=3.7\times10^{-4}$ to the raw infidelity for an atomic temperature of $T=2.7~\mu\mathrm{K}$. In our numerical model, Doppler dephasing is the largest remaining in-subspace contribution, giving a postselected error of $\varepsilon_{\mathrm{ps}}=2.8\times 10^{-4}$.

Finite Rydberg blockade within each addressed dimer provides an additional contribution. The gate pulse is designed in the perfect-blockade limit and is not explicitly compensated for finite interaction strength. At the nominal spacing $R=2.0~\mu\mathrm{m}$, the interaction already starts to deviate from the simple van der Waals regime and enters the crossover toward the F\"orster-interaction regime. To capture this effect, we use a separate $R$-dependent model based on Rydberg-pair eigenstates calculated using the multichannel quantum defect theory (MQDT) model of Ref.~\cite{peper_spectroscopy_2025}. In this model, for each interatomic distance $R$, the pair-state Hamiltonian is written in the MQDT eigenbasis, and the laser couplings are computed from the overlaps with the original product-state basis. At the nominal spacing, this model gives a finite-blockade contribution of $\varepsilon_{\mathrm{raw}}=1.6 \times 10^{-4}$ and $\varepsilon_{\mathrm{ps}}=3 \times 10^{-5}$.

On top of this nominal finite-blockade error, the interatomic spacing also fluctuates shot to shot because of the finite temperature of the atoms. We therefore average the simulated gate fidelity over the thermal distribution of interatomic distances. The radial temperature is inferred from Doppler-sensitive Ramsey measurements, while the axial temperature is constrained by comparing measured Rydberg pair-loss spectra with MQDT-based simulations. This comparison is consistent with using $T=2.7~\mu\mathrm{K}$ in both the radial and axial directions, giving an additional contribution from interatomic-distance fluctuations of $\varepsilon_{\mathrm{raw}}=1.5 \times 10^{-4}$ and $\varepsilon_{\mathrm{ps}}=8 \times 10^{-5}$.

Laser phase noise (PN) contributes a raw error of $\varepsilon_{\mathrm{raw}} = 1.4 \times 10^{-4}$ and a postselected error of $\varepsilon_{\mathrm{ps}} = 1.0 \times 10^{-4}$. Laser amplitude noise (AN) contributes a smaller raw error of $\varepsilon_{\mathrm{raw}} = 1 \times 10^{-5}$, mostly through leakage. We also consider an unwanted $\pi$-polarized component of the laser field, whose relative Rabi frequency is measured in a separate experiment to be $\Omega_\pi/\Omega_{\sigma^-} < 1.8 \times 10^{-3}$. The resulting contribution to the gate infidelity is below $10^{-5}$ and is not shown in Fig.~\ref{fig:fidelity}f.

Adding these contributions gives a total simulated raw infidelity of about $\varepsilon_{\mathrm{raw}} = 4.0 \times 10^{-3}$ and a postselected infidelity of about $\varepsilon_{\mathrm{ps}} = 6.6 \times 10^{-4}$. The simulated raw error agrees well with the measured value, while the measured postselected error remains slightly higher than the model prediction. We expect that this remaining discrepancy is mainly due to the difficulty of quantitatively modeling the full Rydberg interaction involving the unwanted $\ket{r'}$ state. This interaction can introduce additional leakage channels, including population in $\ket{r'r'}$ and in other Rydberg-pair states mixed by the interaction. Since these channels are not fully captured by the model or by the scanned Hessian eigendirections, they are not expected to be completely removed by the optimization. 

%% file: appendix/three_outcome_measurement.tex
\section{Three-Outcome Measurement}\label{sec:Three-outcome}

\begin{figure}
    \centering
    \includegraphics[width=3.375 in]{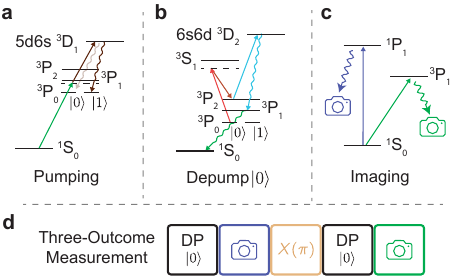}
    \caption{
    \textbf{Three-outcome measurement.}
    (a,b,c) Level scheme and pulse sequence used for state preparation and readout.
    (a) Atoms are initialized in the metastable $6s6p\,^3P_0$ manifold using optical pumping through the $5d6s\,^3D_1$ state.
    (b) State-selective depumping transfers population in $\ket{0}$ from $^3P_0$ to $^3P_2$ using a two-photon Raman transition through $6s7s\,^3S_1$, followed by depumping to $^1S_0$ through $6s6d\,^3D_2$.
    (c) Atoms in $^1S_0$ are detected by fluorescence imaging on either the $^1S_0\leftrightarrow{}^1P_1$ transition at $399~\mathrm{nm}$ or the $^1S_0\leftrightarrow{}^3P_1$ transition at $556~\mathrm{nm}$.
    (d) The three-outcome measurement consists of state-selective depumping of $\ket{0}$, a first image at $399~\mathrm{nm}$, an RF $\pi$ pulse, a second state-selective depumping step, and a final image at $556~\mathrm{nm}$.
    }
    \label{fig:state_selective_readout}
\end{figure}

To initialize the atoms in the metastable $^3P_0$ state, the atoms are optically pumped through a two-photon transition from $^1S_0$ to $^3D_1$ through the intermediate state $^3P_1$, as shown in Fig.~\ref{fig:state_selective_readout}a. The state-selective depumping sequence~\cite{li_fast_2025}, shown in Fig.~\ref{fig:state_selective_readout}b, is used to independently measure the $\ket{0}$ and $\ket{1}$ populations. In the first depumping step, atoms in $\ket{0}$ are transferred to the ground state and detected by destructive fluorescence imaging on the $^1S_0 \, \leftrightarrow\, ^1P_1$ transition at 399~nm (Fig.~\ref{fig:state_selective_readout}c). A subsequent $\pi$ pulse maps the $\ket{1}$ population onto $\ket{0}$, after which the same depumping procedure sends the mapped population to $^1S_0$ for imaging on the $^1S_0 \leftrightarrow {}^3P_1$ transition at 556~nm. Together, the two images define a three-outcome measurement (Fig.~\ref{fig:state_selective_readout}d): atoms detected in the first image are assigned to $\ket{0}$, atoms detected in the second image are assigned to $\ket{1}$, and atoms that are dark in both images are classified as lost.

Depumping from $\ket{0}$ is implemented in two steps. First, a two-photon Raman transition coherently drives the population to the $6s6p\,{}^3P_2$ manifold. A resonant depumping laser then couples this manifold to the $6s6d\,^3D_2$ state, which decays to the ground state via $6s6p\,^3P_1$. The Raman transition is driven through the intermediate state $6s7s\,^3S_1$, $F=1/2$, with a red detuning of $\Delta\simeq 2\pi \times 12~\mathrm{GHz}$. It is resonant with $\ket{0}\rightarrow \ket{6s6p\,^3P_2,F=3/2,m_F=-1/2}$, while the corresponding transition from $\ket{1}$ is detuned by about $12~\mathrm{MHz}$ through the Zeeman splitting of the $6s6p\,^3P_2$ manifold. The 649~nm and 770~nm lasers have single-photon Rabi frequencies $\Omega_{649} = 2\pi \times 71(2)~\mathrm{MHz}$ and $\Omega_{770} = 2\pi \times 130(3)~\mathrm{MHz}$, corresponding to an effective two-photon Raman Rabi frequency of $\Omega_{649+770} = 2\pi \times 0.278(1)~\mathrm{MHz}$.

To avoid inhomogeneous light shifts during gate operations, the optical tweezers are modulated at $400~\mathrm{kHz}$~\cite{zhang_leveraging_2025}. Consequently, the Raman-transfer and depumping steps are synchronized to the trap-off portion of the modulation cycle. Applying the depumping sequence with the traps off also avoids atom loss due to anti-trapping of the $6s6p\,^3P_2$ state. Since the finite trap-off window is shorter than the nominal Raman $\pi$-pulse time, the 550~ns Raman pulse and depumping sequence are repeated 60 times to ensure efficient transfer of the $\ket{0}$ population to the ground state.

To characterize the performance of the three-outcome measurement, we prepare atoms in either $\ket{0}$ or $\ket{1}$ and apply the full measurement sequence. The resulting photon-count distributions are shown in Fig.~\ref{fig:SSR_SQRB}a, and the outcome probabilities are summarized in Table~\ref{tab:ssr_spam_comparison}. The observed infidelities include contributions from both state-preparation and measurement (SPAM) errors.

We find that the spin-flip errors $P_0(1) = 0.17(1)\%$ and $P_1(0) = 0.83(3)\%$ are largely limited by off-resonant scattering of the 649~nm Raman light, which can accidentally depump population in $\ket{1}$ during the $\ket{0}$-selective depumping sequence. We independently measure that $0.63(9)\%$ of atoms prepared in $\ket{1}$ are depumped to $^1S_0$ during state-selective readout, in agreement with a numerical simulation giving $0.6(1)\%$. During readout, this process directly contributes to the misidentification of atoms prepared in $\ket{1}$ as $\ket{0}$, since accidental depumping produces a bright signal in the first image.

The same scattering mechanism also limits the initial spin purity during a separate spin-purification stage. In this stage, atoms are optically pumped from $^1S_0$ into the metastable manifold, with populations in $\ket{0}$ and $\ket{1}$ set by the branching ratios of the pumping sequence as shown in Fig.~\ref{fig:state_selective_readout}a. The $\ket{0}$ population is then selectively depumped back to $^1S_0$, and the cycle is repeated 120 times to accumulate population in $\ket{1}$. However, off-resonant scattering can also depump a small fraction of the desired $\ket{1}$ population back to $^1S_0$. Upon repumping, this population can enter either metastable spin state, producing a finite residual spin impurity before readout. Together, readout-induced misidentification and imperfect spin purification account for the majority of the measured spin-flip errors.

\begin{table}
    \centering
    \begin{tabular}{c c c c}
        \hline
        Initial state & Outcome & Experiment (\%) & Model (\%) \\
        \hline
        \multirow{3}{*}{$\ket{0}$}
            & $P_0(0)$ & 98.75(3) & 98.67(12) \\
            & $P_0(1)$ & 0.17(1)  & 0.14(7) \\
            & $P_0(\mathrm{loss})$ & 1.07(3) & 1.19(9) \\
        \hline
        \multirow{3}{*}{$\ket{1}$}
            & $P_1(0)$ & 0.83(3)  & 0.70(10) \\
            & $P_1(1)$ & 97.53(5) & 97.85(15) \\
            & $P_1(\mathrm{loss})$ & 1.65(4) & 1.45(13) \\
        \hline
    \end{tabular}
    \caption{\textbf{Comparison between the measured and modeled SPAM errors for the three-outcome state-selective readout.}}
    \label{tab:ssr_spam_comparison}
\end{table}

To quantify the total SPAM infidelity, we model the full sequence using a four-state transition-matrix model that tracks the atom population in $\ket{1}$, $\ket{0}$, $^1S_0$, and loss. Each elementary operation is represented by a transition matrix between these states, with transition probabilities determined from independent calibration measurements. The full model is obtained by composing the matrices for the successive steps of the sequence, yielding the expected probabilities for the three measurement outcomes. The calibrated error channels include loss during optical pumping and depumping, $0.005(1)$, whose microscopic origin has not yet been unambiguously identified, as well as image-classification errors for both fluorescence images. We denote false-positive errors, in which a dark atom is classified as bright, by $\epsilon_{\rm FP}$, and false-negative errors, in which a bright atom is classified as dark, by $\epsilon_{\rm FN}$. For the destructive 399~nm image, we measure $\epsilon^{399}_{\rm FP}=0.0020(16)$ and $\epsilon^{399}_{\rm FN}=0.00090(21)$. For each 556~nm image, we measure $\epsilon^{556}_{\rm FP}=3\times10^{-4}$, $\epsilon^{556}_{\rm FN}=2\times10^{-4}$, and an imaging-induced atom-loss probability of $0.004(1)$. The model also accounts for the finite duration of each operation by including decay from the metastable $^3P_0$ manifold, with measured lifetime $1.50(7)$~s, and atom loss from the tweezer, with lifetime $25(5)$~s.

A comparison between the measured and predicted probabilities is shown in Table~\ref{tab:ssr_spam_comparison}. Overall, the measured infidelities are consistent with the values predicted by the model, indicating that the dominant SPAM error mechanisms are captured. The three-outcome measurement is mainly limited by Raman-beam scattering and by errors accumulated over the finite sequence duration. Raman-beam scattering produces spin-changing SPAM errors through off-resonant depumping of the nominally dark spin state, while the finite sequence duration leads to metastable-state decay and atom loss. These mechanisms suggest several routes for improvement. Scattering-induced spin flips could be suppressed by increasing the available Raman laser power, allowing operation farther from the intermediate-state resonance while maintaining the same Raman Rabi frequency. The $^3P_0$ lifetime, currently limited in part by scattering from the 488~nm trapping light~\cite{ma_high-fidelity_2023}, could also be extended by using a trapping wavelength with a lower scattering rate, such as 780~nm. Finally, the sequence duration could be reduced by increasing the RF Rabi frequency beyond its present value of approximately $300~\mathrm{Hz}$, directly addressing the $\ket{1}\rightarrow\ket{6s6p\,^3P_2}$ Raman transition in the second depumping step to eliminate the RF $\pi$ pulse, or operating at a constant magnetic field to avoid ramping stages in the protocol~\cite{li_fast_2025}.

%% file: appendix/rb_sequence.tex
\begin{figure}[b!]
    \centering
    \includegraphics[]{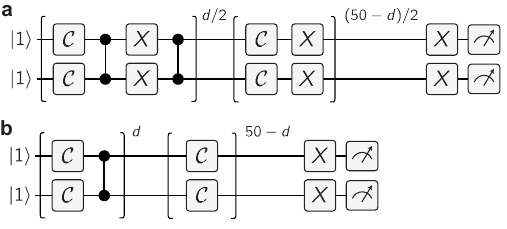}
    \caption{\textbf{Global randomized benchmarking circuits.} 
    (a) Circuit used for echoed global randomized benchmarking. 
    (b) Circuit used for non-echoed global randomized benchmarking, which is sensitive to single-qubit phase errors. After each CZ gate, an unshown virtual ($Z$) rotation is applied by shifting the phases of all subsequent gates.
    }
    \label{fig:rb}
\end{figure}

\section{Randomized benchmarking of CZ gates}

For most of this work, we characterize the CZ gates using the echoed global randomized benchmarking sequence of Ref.~\cite{evered_high-fidelity_2023}, which cancels sensitivity to single-qubit phase errors while preserving the number of single-qubit operations in the sequence (Fig.~\ref{fig:rb}a). The only exception is Fig.~\ref{fig:Fig3}d, where we instead use a global randomized benchmarking sequence similar to that of Ref.~\cite{peper_spectroscopy_2025}, without the echo cancellation (Fig.~\ref{fig:rb}b). This latter sequence remains sensitive to single-qubit phase errors, which we leverage to more accurately measure the fidelity response along the Hessian-sensitive directions.

%% file: ref.bib
@article{glaser_closed-loop_2025,
  title= {Closed-loop optimization for high-fidelity controlled-$Z$ gates in superconducting qubits},
  author= {Glaser, N.J. and Roy, F.A. and Tsitsilin, I. and Koch, L. and Bruckmoser, N. and Schirk, J. and Romeiro, J.H. and Huber, G.B.P. and Wallner, F. and Singh, M. and Krylov, G. and Marx, A. and S\"odergren, L. and Schneider, C.M.F. and Werninghaus, M. and Filipp, S.},
  journal= {Phys. Rev. Appl.},
  volume= {24},
  issue= {2},
  pages= {024048},
  numpages= {15},
  year= {2025},
  month= {Aug},
  publisher= {American Physical Society},
  doi= {10.1103/pckq-2csc},
  url= {https://link.aps.org/doi/10.1103/pckq-2csc}
}

@article{sporl_optimal_2007,
  title= {Optimal control of coupled Josephson qubits},
  author= {Sp\"orl, A. and Schulte-Herbr\"uggen, T. and Glaser, S. J. and Bergholm, V. and Storcz, M. J. and Ferber, J. and Wilhelm, F. K.},
  journal= {Physical Review A},
  volume= {75},
  issue= {1},
  pages= {012302},
  numpages= {9},
  year= {2007},
  month= {Jan},
  publisher= {American Physical Society},
  doi= {10.1103/PhysRevA.75.012302},
  url= {https://link.aps.org/doi/10.1103/PhysRevA.75.012302}
}

@article{kang_batch_2021,
  title= {Batch Optimization of Frequency-Modulated Pulses for Robust Two-Qubit Gates in Ion Chains},
  author= {Kang, Mingyu and Liang, Qiyao and Zhang, Bichen and Huang, Shilin and Wang, Ye and Fang, Chao and Kim, Jungsang and Brown, Kenneth R.},
  journal= {Phys. Rev. Appl.},
  volume= {16},
  issue= {2},
  pages= {024039},
  numpages= {11},
  year= {2021},
  month= {Aug},
  publisher= {American Physical Society},
  doi= {10.1103/PhysRevApplied.16.024039},
  url= {https://link.aps.org/doi/10.1103/PhysRevApplied.16.024039}
}

@article{leung_robust_2018,
  title= {Robust 2-Qubit Gates in a Linear Ion Crystal Using a Frequency-Modulated Driving Force},
  author= {Leung, Pak Hong and Landsman, Kevin A. and Figgatt, Caroline and Linke, Norbert M. and Monroe, Christopher and Brown, Kenneth R.},
  journal= {Physical Review Letters},
  volume= {120},
  issue= {2},
  pages= {020501},
  numpages= {4},
  year= {2018},
  month= {Jan},
  publisher= {American Physical Society},
  doi= {10.1103/PhysRevLett.120.020501},
  url= {https://link.aps.org/doi/10.1103/PhysRevLett.120.020501}
}

@article{choi_optimal_2014,
  title= {Optimal Quantum Control of Multimode Couplings between Trapped Ion Qubits for Scalable Entanglement},
  author= {Choi, T. and Debnath, S. and Manning, T. A. and Figgatt, C. and Gong, Z.-X. and Duan, L.-M. and Monroe, C.},
  journal= {Physical Review Letters},
  volume= {112},
  issue= {19},
  pages= {190502},
  numpages= {5},
  year= {2014},
  month= {May},
  publisher= {American Physical Society},
  doi= {10.1103/PhysRevLett.112.190502},
  url= {https://link.aps.org/doi/10.1103/PhysRevLett.112.190502}
}

@article{levine_parallel_2019,
    title= {Parallel {Implementation} of {High}-{Fidelity} {Multiqubit} {Gates} with {Neutral} {Atoms}},
    volume= {123},
    url= {https://link.aps.org/doi/10.1103/PhysRevLett.123.170503},
    doi= {10.1103/PhysRevLett.123.170503},
    number= {17},
    journal= {Physical Review Letters},
    publisher= {American Physical Society},
    author= {Levine, Harry and Keesling, Alexander and Semeghini, Giulia and Omran, Ahmed and Wang, Tout T. and Ebadi, Sepehr and Bernien, Hannes and Greiner, Markus and Vuletić, Vladan and Pichler, Hannes and Lukin, Mikhail D.},
    month= oct,
    year= {2019},
    pages= {170503},
}

@article{jandura_time-optimal_2022,
    title= {Time-{Optimal} {Two}- and {Three}-{Qubit} {Gates} for {Rydberg} {Atoms}},
    volume= {6},
    url= {https://quantum-journal.org/papers/q-2022-05-13-712/},
    doi= {10.22331/q-2022-05-13-712},
    journal= {Quantum},
    author= {Jandura, Sven and Pupillo, Guido},
    month= may,
    year= {2022},
    pages= {712},
}

@article{jandura_optimizing_2023,
    title= {Optimizing {Rydberg} {Gates} for {Logical}-{Qubit} {Performance}},
    volume= {4},
    url= {https://link.aps.org/doi/10.1103/PRXQuantum.4.020336},
    doi= {10.1103/PRXQuantum.4.020336},
    number= {2},
    journal= {PRX Quantum},
    publisher= {American Physical Society},
    author= {Jandura, Sven and Thompson, Jeff D. and Pupillo, Guido},
    month= jun,
    year= {2023},
    pages= {020336},
}

@article{fromonteil_protocols_2023,
    title= {Protocols for {Rydberg} {Entangling} {Gates} {Featuring} {Robustness} against {Quasistatic} {Errors}},
    volume= {4},
    url= {https://link.aps.org/doi/10.1103/PRXQuantum.4.020335},
    doi= {10.1103/PRXQuantum.4.020335},
    number= {2},
    journal= {PRX Quantum},
    publisher= {American Physical Society},
    author= {Fromonteil, Charles and Bluvstein, Dolev and Pichler, Hannes},
    month= jun,
    year= {2023},
    pages= {020335},
}

@article{ma_high-fidelity_2023,
    title= {High-fidelity gates and mid-circuit erasure conversion in an atomic qubit},
    volume= {622},
    url= {https://www.nature.com/articles/s41586-023-06438-1},
    doi= {10.1038/s41586-023-06438-1},
    number= {7982},
    journal= {Nature},
    publisher= {Nature Publishing Group},
    author= {Ma, Shuo and Liu, Genyue and Peng, Pai and Zhang, Bichen and Jandura, Sven and Claes, Jahan and Burgers, Alex P. and Pupillo, Guido and Puri, Shruti and Thompson, Jeff D.},
    month= oct,
    year= {2023},
    pages= {279--284},
}

@unpublished{zhang_leveraging_2025,
    title= {Leveraging erasure errors in logical qubits with metastable $^{171}${Yb} atoms},
    url= {http://arxiv.org/abs/2506.13724},
    doi= {10.48550/arXiv.2506.13724},
    publisher= {arXiv},
    author= {Zhang, Bichen and Liu, Genyue and Bornet, Guillaume and Horvath, Sebastian P. and Peng, Pai and Ma, Shuo and Huang, Shilin and Puri, Shruti and Thompson, Jeff D.},
    month= jun,
    year= {2025},
    note= {arXiv:2506.13724},
}

@article{evered_high-fidelity_2023,
    title= {High-fidelity parallel entangling gates on a neutral-atom quantum computer},
    volume= {622},
    url= {https://www.nature.com/articles/s41586-023-06481-y},
    doi= {10.1038/s41586-023-06481-y},
    number= {7982},
    journal= {Nature},
    publisher= {Nature Publishing Group},
    author= {Evered, Simon J. and Bluvstein, Dolev and Kalinowski, Marcin and Ebadi, Sepehr and Manovitz, Tom and Zhou, Hengyun and Li, Sophie H. and Geim, Alexandra A. and Wang, Tout T. and Maskara, Nishad and Levine, Harry and Semeghini, Giulia and Greiner, Markus and Vuletić, Vladan and Lukin, Mikhail D.},
    month= oct,
    year= {2023},
    pages= {268--272},
}

@article{muniz_high-fidelity_2025,
    title= {High-{Fidelity} {Universal} {Gates} in the ${}^{171}$$\mathrm{{Yb}}$ {Ground}-{State} {Nuclear}-{Spin} {Qubit}},
    volume= {6},
    url= {https://link.aps.org/doi/10.1103/PRXQuantum.6.020334},
    doi= {10.1103/PRXQuantum.6.020334},
    number= {2},
    journal= {PRX Quantum},
    publisher= {American Physical Society},
    author= {Muniz, J. A. and Stone, M. and Stack, D. T. and Jaffe, M. and Kindem, J. M. and Wadleigh, L. and Zalys-Geller, E. and Zhang, X. and Chen, C.-A. and Norcia, M. A. and Epstein, J. and Halperin, E. and Hummel, F. and Wilkason, T. and Li, M. and Barnes, K. and Battaglino, P. and Bohdanowicz, T. C. and Booth, G. and Brown, A. and Brown, M. O. and Cairncross, W. B. and Cassella, K. and Coxe, R. and Crow, D. and Feldkamp, M. and Griger, C. and Heinz, A. and Jones, A. M. W. and Kim, H. and King, J. and Kotru, K. and Lauigan, J. and Marjanovic, J. and Megidish, E. and Meredith, M. and McDonald, M. and Morshead, R. and Narayanaswami, S. and Nishiguchi, C. and Paule, T. and Pawlak, K. A. and Pudenz, K. L. and P\'erez, D. Rodr\'iguez and Ryou, A. and Simon, J. and Smull, A. and Urbanek, M. and van de Veerdonk, R. J. M. and Vendeiro, Z. and Wu, T.-Y. and Xie, X. and Bloom, B. J.},
    month= may,
    year= {2025},
    pages= {020334},
}

@article{peper_spectroscopy_2025,
    title= {Spectroscopy and {Modeling} of $^{171}\mathrm{{Yb}}$ {Rydberg} {States} for {High}-{Fidelity} {Two}-{Qubit} {Gates}},
    volume= {15},
    url= {https://link.aps.org/doi/10.1103/PhysRevX.15.011009},
    doi= {10.1103/PhysRevX.15.011009},
    number= {1},
    journal= {Physical Review X},
    publisher= {American Physical Society},
    author= {Peper, Michael and Li, Yiyi and Knapp, Daniel Y. and Bileska, Mila and Ma, Shuo and Liu, Genyue and Peng, Pai and Zhang, Bichen and Horvath, Sebastian P. and Burgers, Alex P. and Thompson, Jeff D.},
    month= jan,
    year= {2025},
    pages= {011009},
}

@article{wu_erasure_2022,
    title= {Erasure conversion for fault-tolerant quantum computing in alkaline earth {Rydberg} atom arrays},
    volume= {13},
    url= {https://www.nature.com/articles/s41467-022-32094-6},
    doi= {10.1038/s41467-022-32094-6},
    number= {1},
    journal= {Nature Communications},
    publisher= {Nature Publishing Group},
    author= {Wu, Yue and Kolkowitz, Shimon and Puri, Shruti and Thompson, Jeff D.},
    month= aug,
    year= {2022},
    pages= {4657},
}

@article{baranes_leveraging_2026,
    title= {Leveraging {Qubit} {Loss} {Detection} in {Fault}-{Tolerant} {Quantum} {Algorithms}},
    volume= {16},
    url= {https://link.aps.org/doi/10.1103/ycwc-3myc},
    doi= {10.1103/ycwc-3myc},
    number= {1},
    journal= {Physical Review X},
    publisher= {American Physical Society},
    author= {Baranes, Gefen and Cain, Madelyn and Ataides, J. Pablo Bonilla and Bluvstein, Dolev and Sinclair, Josiah and Vuletić, Vladan and Zhou, Hengyun and Lukin, Mikhail D.},
    month= jan,
    year= {2026},
    pages= {011002},
}

@article{lis_midcircuit_2023,
    title= {Midcircuit {Operations} {Using} the omg {Architecture} in {Neutral} {Atom} {Arrays}},
    volume= {13},
    url= {https://link.aps.org/doi/10.1103/PhysRevX.13.041035},
    doi= {10.1103/PhysRevX.13.041035},
    number= {4},
    journal= {Physical Review X},
    publisher= {American Physical Society},
    author= {Lis, Joanna W. and Senoo, Aruku and McGrew, William F. and R\"onchen, Felix and Jenkins, Alec and Kaufman, Adam M.},
    month= nov,
    year= {2023},
    pages= {041035},
}

@unpublished{li_fast_2025,
    title= {Fast, continuous and coherent atom replacement in a neutral atom qubit array},
    url= {http://arxiv.org/abs/2506.15633},
    doi= {10.48550/arXiv.2506.15633},
    publisher= {arXiv},
    author= {Li, Yiyi and Bao, Yicheng and Peper, Michael and Li, Chenyuan and Thompson, Jeff D.},
    month= jun,
    year= {2025},
    note= {arXiv:2506.15633},
}

@article{khaneja_optimal_2005,
    title= {Optimal control of coupled spin dynamics: design of {NMR} pulse sequences by gradient ascent algorithms},
    volume= {172},
    shorttitle= {Optimal control of coupled spin dynamics},
    url= {https://www.sciencedirect.com/science/article/pii/S1090780704003696},
    doi= {10.1016/j.jmr.2004.11.004},
    number= {2},
    journal= {Journal of Magnetic Resonance},
    author= {Khaneja, Navin and Reiss, Timo and Kehlet, Cindie and Schulte-Herbr\"uggen, Thomas and Glaser, Steffen J.},
    month= feb,
    year= {2005},
    pages= {296--305},
}

@article{caneva_chopped_2011,
    title= {Chopped random-basis quantum optimization},
    volume= {84},
    url= {https://link.aps.org/doi/10.1103/PhysRevA.84.022326},
    doi= {10.1103/PhysRevA.84.022326},
    number= {2},
    journal= {Physical Review A},
    publisher= {American Physical Society},
    author= {Caneva, Tommaso and Calarco, Tommaso and Montangero, Simone},
    month= aug,
    year= {2011},
    pages= {022326},
}

@article{machnes_tunable_2018,
    title= {Tunable, {Flexible}, and {Efficient} {Optimization} of {Control} {Pulses} for {Practical} {Qubits}},
    volume= {120},
    url= {https://link.aps.org/doi/10.1103/PhysRevLett.120.150401},
    doi= {10.1103/PhysRevLett.120.150401},
    number= {15},
    journal= {Physical Review Letters},
    publisher= {American Physical Society},
    author= {Machnes, Shai and Ass\'emat, Elie and Tannor, David and Wilhelm, Frank K.},
    month= apr,
    year= {2018},
    pages= {150401},
}

@article{palao_quantum_2002,
    title= {Quantum {Computing} by an {Optimal} {Control} {Algorithm} for {Unitary} {Transformations}},
    volume= {89},
    url= {https://link.aps.org/doi/10.1103/PhysRevLett.89.188301},
    doi= {10.1103/PhysRevLett.89.188301},
    number= {18},
    journal= {Physical Review Letters},
    publisher= {American Physical Society},
    author= {Palao, Jos\'e P. and Kosloff, Ronnie},
    month= oct,
    year= {2002},
    pages= {188301},
}

@unpublished{wilhelm_introduction_2020,
    title= {An introduction into optimal control for quantum technologies},
    url= {http://arxiv.org/abs/2003.10132},
    doi= {10.48550/arXiv.2003.10132},
    publisher= {arXiv},
    author= {Wilhelm, Frank K. and Kirchhoff, Susanna and Machnes, Shai and Wittler, Nicolas and Sugny, Dominique},
    month= mar,
    year= {2020},
    note= {arXiv:2003.10132},
}

@article{white_extracting_2004,
doi= {10.1088/0953-4075/37/24/L02},
url= {https://doi.org/10.1088/0953-4075/37/24/L02},
year= {2004},
month= {dec},
publisher= {},
volume= {37},
number= {24},
pages= {L399},
author= {White, J L and Pearson, B J and Bucksbaum, P H},
title= {Extracting quantum dynamics from genetic learning algorithms through principal control analysis},
journal= {Journal of Physics B: Atomic, Molecular and Optical Physics},
}

@article{sivak_model-free_2022,
    title= {Model-{Free} {Quantum} {Control} with {Reinforcement} {Learning}},
    volume= {12},
    url= {https://link.aps.org/doi/10.1103/PhysRevX.12.011059},
    doi= {10.1103/PhysRevX.12.011059},
    number= {1},
    journal= {Physical Review X},
    publisher= {American Physical Society},
    author= {Sivak, V. V. and Eickbusch, A. and Liu, H. and Royer, B. and Tsioutsios, I. and Devoret, M. H.},
    month= mar,
    year= {2022},
    pages= {011059},
}

@article{kelly_optimal_2014,
    title= {Optimal {Quantum} {Control} {Using} {Randomized} {Benchmarking}},
    volume= {112},
    url= {https://link.aps.org/doi/10.1103/PhysRevLett.112.240504},
    doi= {10.1103/PhysRevLett.112.240504},
    number= {24},
    journal= {Physical Review Letters},
    publisher= {American Physical Society},
    author= {Kelly, J. and Barends, R. and Campbell, B. and Chen, Y. and Chen, Z. and Chiaro, B. and Dunsworth, A. and Fowler, A. G. and Hoi, I.-C. and Jeffrey, E. and Megrant, A. and Mutus, J. and Neill, C. and O'Malley, P. J. J. and Quintana, C. and Roushan, P. and Sank, D. and Vainsencher, A. and Wenner, J. and White, T. C. and Cleland, A. N. and Martinis, John M.},
    month= jun,
    year= {2014},
    pages= {240504},
}

@article{werninghaus_leakage_2021,
    title= {Leakage reduction in fast superconducting qubit gates via optimal control},
    volume= {7},
    url= {https://www.nature.com/articles/s41534-020-00346-2},
    doi= {10.1038/s41534-020-00346-2},
    number= {1},
    journal= {npj Quantum Information},
    publisher= {Nature Publishing Group},
    author= {Werninghaus, M. and Egger, D. J. and Roy, F. and Machnes, S. and Wilhelm, F. K. and Filipp, S.},
    month= jan,
    year= {2021},
    pages= {14},
}

@article{porotti_gradient-ascent_2023,
    title= {Gradient-{Ascent} {Pulse} {Engineering} with {Feedback}},
    volume= {4},
    url= {https://link.aps.org/doi/10.1103/PRXQuantum.4.030305},
    doi= {10.1103/PRXQuantum.4.030305},
    number= {3},
    journal= {PRX Quantum},
    publisher= {American Physical Society},
    author= {Porotti, Riccardo and Peano, Vittorio and Marquardt, Florian},
    month= jul,
    year= {2023},
    pages= {030305},
}

@article{theis_high-fidelity_2016,
    title= {High-fidelity {Rydberg}-blockade entangling gate using shaped, analytic pulses},
    volume= {94},
    url= {https://doi.org/10.1103/physreva.94.032306},
    doi= {10.1103/physreva.94.032306},
    number= {3},
    journal= {Physical Review A},
    author= {Theis, L. S. and Motzoi, F. and Wilhelm, F. K. and Saffman, M.},
    month= {sep},
    year= {2016},
    pages= {032306},
}

@article{Saskin.2019,
  title= {Narrow-Line Cooling and Imaging of Ytterbium Atoms in an Optical Tweezer Array},
  author= {Saskin, S. and Wilson, J. T. and Grinkemeyer, B. and Thompson, J. D.},
  journal= {Physical Review Letters},
  volume= {122},
  issue= {14},
  pages= {143002},
  numpages= {6},
  year= {2019},
  month= {Apr},
  publisher= {American Physical Society},
  doi= {10.1103/PhysRevLett.122.143002},
  url= {https://link.aps.org/doi/10.1103/PhysRevLett.122.143002}
}

@article{rabitz_topology_2006,
    title= {Topology of optimally controlled quantum mechanical transition probability landscapes},
    volume= {74},
    url= {https://link.aps.org/doi/10.1103/PhysRevA.74.012721},
    doi= {10.1103/PhysRevA.74.012721},
    number= {1},
    journal= {Physical Review A},
    publisher= {American Physical Society},
    author= {Rabitz, H. and Ho, T.-S. and Hsieh, M. and Kosut, R. and Demiralp, M.},
    month= jul,
    year= {2006},
    pages= {012721},
}

@article{shen_quantum_2006,
    title= {Quantum optimal control: {Hessian} analysis of the control landscape},
    volume= {124},
    shorttitle= {Quantum optimal control},
    url= {https://doi.org/10.1063/1.2198836},
    doi= {10.1063/1.2198836},
    number= {20},
    journal= {The Journal of Chemical Physics},
    author= {Shen, Zhenwen and Hsieh, Michael and Rabitz, Herschel},
    month= may,
    year= {2006},
    pages= {204106},
}

@article{Jaksch.2000, 
year= {2000}, 
title= {{Fast Quantum Gates for Neutral Atoms}}, 
author= {Jaksch, D. and Cirac, J. I. and Zoller, P. and Rolston, S. L. and C\^ot\'e, R. and Lukin, M. D.}, 
journal= {Physical Review Letters}, 
doi= {10.1103/physrevlett.85.2208}, 
pmid= {10970499}, 
pages= {2208-2211}, 
number= {10}, 
volume= {85},
month= {sep}, 
url= {https://doi.org/10.1103/physrevlett.85.2208},
}

@article{yu2026taming,
  title={Taming Rydberg Decay with Measurement-Based Quantum Computation},
  author={Yu, Cheng-Cheng and Chen, Zi-Han and Deng, Yu-Hao and Lu, Chao-Yang and Chen, Ming-Cheng and Pan, Jian-Wei},
  journal={Physical Review Letters},
  volume={136},
  number={16},
  pages={160601},
  year={2026},
  publisher={APS}
}

@article{sahay2023high,
  title={High-threshold codes for neutral-atom qubits with biased erasure errors},
  author={Sahay, Kaavya and Jin, Junlan and Claes, Jahan and Thompson, Jeff D and Puri, Shruti},
  journal={Physical Review X},
  volume={13},
  number={4},
  pages={041013},
  year={2023},
  publisher={APS}
}

@article{norcia2023midcircuit,
  title={Midcircuit qubit measurement and rearrangement in a $^{171}${Yb}  atomic array},
  author={Norcia, MA and Cairncross, WB and Barnes, K and Battaglino, P and Brown, A and Brown, MO and Cassella, K and Chen, C-A and Coxe, R and Crow, D and others},
  journal={Physical Review X},
  volume={13},
  number={4},
  pages={041034},
  year={2023},
  publisher={APS}
}

@article{chow2024circuit,
  title={Circuit-based leakage-to-erasure conversion in a neutral-atom quantum processor},
  author={Chow, Matthew NH and Buchemmavari, Vikas and Omanakuttan, Sivaprasad and Little, Bethany J and Pandey, Saurabh and Deutsch, Ivan H and Jau, Yuan-Yu},
  journal={PRX Quantum},
  volume={5},
  number={4},
  pages={040343},
  year={2024},
  publisher={APS}
}

@article{evered2026high,
  title={High-fidelity entangling gates and nonlocal circuits with neutral atoms},
  author={Evered, Simon J and Xu, Muqing and Li, Sophie H and Geim, Alexandra A and Ataides, J and Kalinowski, Marcin and Bluvstein, Dolev and Maskara, Nishad and Kokail, Christian and Greiner, Markus and others},
  journal={arXiv:2604.25987},
  year={2026}
}

@article{isenhower2010demonstration,
  title={Demonstration of a neutral atom controlled-NOT quantum gate},
  author={Isenhower, L and Urban, E and Zhang, XL and Gill, AT and Henage, T and Johnson, Todd A and Walker, TG and Saffman, M},
  journal={Physical Review Letters},
  volume={104},
  number={1},
  pages={010503},
  year={2010},
  publisher={APS}
}

@article{wilk2010entanglement,
  title={Entanglement of two individual neutral atoms using Rydberg blockade},
  author={Wilk, Tatjana and Ga{\"e}tan, A and Evellin, C and Wolters, J and Miroshnychenko, Y and Grangier, P and Browaeys, Antoine},
  journal={Physical Review Letters},
  volume={104},
  number={1},
  pages={010502},
  year={2010},
  publisher={APS}
}

@article{wilson_trapping_2022,
    title= {Trapping {Alkaline} {Earth} {Rydberg} {Atoms} {Optical} {Tweezer} {Arrays}},
    volume= {128},
    url= {https://link.aps.org/doi/10.1103/PhysRevLett.128.033201},
    doi= {10.1103/PhysRevLett.128.033201},
    number= {3},
    journal= {Physical Review Letters},
    publisher= {American Physical Society},
    author= {Wilson, J.T. and Saskin, S. and Meng, Y. and Ma, S. and Dilip, R. and Burgers, A.P. and Thompson, J.D.},
    month= jan,
    year= {2022},
    pages= {033201},
}

@article{saffman2020symmetric,
  title={Symmetric Rydberg controlled-Z gates with adiabatic pulses},
  author={Saffman, M and Beterov, II and Dalal, A and P{\'a}ez, EJ and Sanders, BC},
  journal={Physical Review A},
  volume={101},
  number={6},
  pages={062309},
  year={2020},
  publisher={APS}
}

@article{ho_landscape_2009,
    title= {Landscape of unitary transformations in controlled quantum dynamics},
    volume= {79},
    url= {https://link.aps.org/doi/10.1103/PhysRevA.79.013422},
    doi= {10.1103/PhysRevA.79.013422},
    number= {1},
    journal= {Physical Review A},
    publisher= {American Physical Society},
    author= {Ho, Tak-San and Dominy, Jason and Rabitz, Herschel},
    month= jan,
    year= {2009},
    pages= {013422},
}

@unpublished{berger_dimensionality_2024,
    title= {Dimensionality reduction for closed-loop quantum gate calibration},
    url= {http://arxiv.org/abs/2412.05230},
    doi= {10.48550/arXiv.2412.05230},
    publisher= {arXiv},
    author= {Berger, Emma and Maurya, Vivek and McIntyre, Z. M. and Wei, Ken Xuan and Haas, Holger and Puzzuoli, Daniel},
    month= dec,
    year= {2024},
    note= {arXiv:2412.05230},
}

@article{pagano_error_2022,
    title= {Error budgeting for a controlled-phase gate with strontium-88 {Rydberg} atoms},
    volume= {4},
    url= {https://link.aps.org/doi/10.1103/PhysRevResearch.4.033019},
    doi= {10.1103/PhysRevResearch.4.033019},
    number= {3},
    journal= {Physical Review Research},
    publisher= {American Physical Society},
    author= {Pagano, Alice and Weber, Sebastian and Jaschke, Daniel and Pfau, Tilman and Meinert, Florian and Montangero, Simone and B\"uchler, Hans Peter},
    month= jul,
    year= {2022},
    pages= {033019},
}

@article{bluvstein_fault-tolerant_2026,
    title= {A fault-tolerant neutral-atom architecture for universal quantum computation},
    volume= {649},
    url= {https://www.nature.com/articles/s41586-025-09848-5},
    doi= {10.1038/s41586-025-09848-5},
    number= {8095},
    journal= {Nature},
    publisher= {Nature Publishing Group},
    author= {Bluvstein, Dolev and Geim, Alexandra A. and Li, Sophie H. and Evered, Simon J. and Bonilla Ataides, J. Pablo and Baranes, Gefen and Gu, Andi and Manovitz, Tom and Xu, Muqing and Kalinowski, Marcin and Majidy, Shayan and Kokail, Christian and Maskara, Nishad and Trapp, Elias C. and Stewart, Luke M. and Hollerith, Simon and Zhou, Hengyun and Gullans, Michael J. and Yelin, Susanne F. and Greiner, Markus and Vuletić, Vladan and Cain, Madelyn and Lukin, Mikhail D.},
    month= jan,
    year= {2026},
    pages= {39--46},
}

@article{perrin_quantum_2025,
    title= {Quantum {Error} {Correction} resilient against {Atom} {Loss}},
    volume= {9},
    url= {https://quantum-journal.org/papers/q-2025-10-13-1884/},
    doi= {10.22331/q-2025-10-13-1884},
    journal= {Quantum},
    publisher= {Verein zur F\"orderung des Open Access Publizierens in den Quantenwissenschaften},
    author= {Perrin, Hugo and Jandura, Sven and Pupillo, Guido},
    month= oct,
    year= {2025},
    pages= {1884},
}

@article{rol_time-domain_2020,
    title= {Time-domain characterization and correction of on-chip distortion of control pulses in a quantum processor},
    volume= {116},
    url= {https://doi.org/10.1063/1.5133894},
    doi= {10.1063/1.5133894},
    number= {5},
    journal= {Applied Physics Letters},
    author= {Rol, M. A. and Ciorciaro, L. and Malinowski, F. K. and Tarasinski, B. M. and Sagastizabal, R. E. and Bultink, C. C. and Salathe, Y. and Haandbaek, N. and Sedivy, J. and DiCarlo, L.},
    month= feb,
    year= {2020},
    pages= {054001},
}

@article{hellings_calibrating_2025,
    title= {Calibrating {Magnetic} {Flux} {Control} in {Superconducting} {Circuits} by {Compensating} {Distortions} on {Time} {Scales} from {Nanoseconds} up to {Tens} of {Microseconds}},
    url= {https://doi.org/10.1103/1qhb-r4fb},
    journal= {Physical Review Research},
    author= {Hellings, Christoph and Lacroix, Nathan and Remm, Ants and Boell, Richard and Herrmann, Johannes and Lazăr, Stefania and Krinner, Sebastian and Swiadek, Fran\c{c}ois and Andersen, Christian Kraglund and Eichler, Christopher and Wallraff, Andreas},
    month= {nov},
    year= {2025},
    note= {arXiv:2503.04610},
    volume= {7},
    number= {4},
    doi= {10.1103/1qhb-r4fb},
}
